\documentclass{pinchcr}   
\usepackage{amsmath,amssymb}

\title{The few-atom problem}

\author{D.S. Petrov}
\affiliation{Laboratoire de Physique Th\'{e}orique et Mod\`{e}les Statistiques, CNRS,
Universit\'{e} Paris Sud, 91405 Orsay, France \\
Russian Research Center Kurchatov Institute,
Kurchatov Square, 123182 Moscow, Russia}

\begin{document}

\maketitle

\preface

Few-body problem plays a very important role in the physics of ultra-cold gases. It describes the structure of molecules and their formation in three-body collisions (three-body recombination), atom-molecule and molecule-molecule collisional properties, structure of trimers and larger clusters, numerous phenomena which go under the name of Efimov physics, and many other problems.

In ultracold gases we benefit from a remarkable separation of scales. The atomic de Broglie wavelengths are much larger than the range, $R_e$, of the van der Waals interatomic potential. The ultracold regime is defined by the inequality $kR_e\ll 1$, where $k$ is the typical atomic momentum. In this regime very few terms in the effective range expansion for the scattering amplitude suffice, and most of the time the interaction is characterized by a single parameter -- the $s$-wave scattering length $a$. From the two-body viewpoint all short-range potentials are equivalent as long as they have the same scattering length, and, therefore, one can use an idealized zero-range potential (pseudopotential) with the same $a$. The zero-range model well describes the weakly interacting BEC (Huang 1963) as well as the whole range of the BCS-BEC crossover in fermionic mixtures (Giorgini {\it et al.} 2008).

There is a class of few-body problems where both long and short lengthscales are important. For example, the knowledge of $a$ is not sufficient for calculating the spectrum of Efimov trimers or the rates of recombination and relaxation to deeply bound molecular states. Such problems can be solved in the {\it universal} limit, $|a|\gg R_e$, by introducing the so-called three-body parameter, which absorbs all the short-range three-body physics in the same manner as the scattering length absorbs the short-range two-body physics. Universality (Braaten and Hammer 2006, 2007) in this context reflects the amazing fact that different systems with the same scattering length and three-body parameter exhibit the same physics. The possibility to modify $a$ in atomic gases by using Feshbach resonances makes ultracold gases an ideal playground to check the universal theory. The Efimov effect predicted 40 years ago (Efimov 1970) has been first observed in a cold gas of $^{133}$Cs in Innsbruck (Kraemer {\it et al.} 2006, see also Ferlaino {\it et al.} 2011 for review) and subsequently in other alkali atoms and mixtures (Ottenstein {\it et al.} 2008; Huckans {\it et al.} 2009; Zaccanti {\it et al.} 2009; Barontini {\it et al.} 2009; Gross {\it et al.} 2009, 2010; Pollack {\it et al.} 2009; Nakajima {\it et al.} 2010, 2011; Lompe {\it et al.} 2010). 

The few-body analysis, apart from being interesting on its own, can be used in many-body problems to integrate out few-body degrees of freedom, thus making the many-body problem more tractable. For example, a two-component fermionic mixture on the BEC side of the BCS-BEC crossover is actually a Bose gas of molecules (or an atom-molecule mixture in the density imbalanced case). If the molecules are sufficiently small ($a$ is much smaller than the interparticle distance), we can forget about their composite nature and treat them as elementary objects as long as the atom-molecule and molecule-molecule scattering parameters are known, the latter being given by the solution of the corresponding three- and four-atom problems. Here we profit from the fact that the few- and many-body processes separate on the length and energy scales. Then the equation of state and the stability of the system with respect to collapse or phase separation are much easier to analyze. We can also mention lattice problems where in order to describe the tight binding limit one has to solve the few-body problem on a single site and use the result as an input for the Hubbard model.  

Another path from ``few'' to ``many'' is the high-temperature virial expansion, which consists of solving few-body problems with more and more particles (Huang 1963; Liu {\it et al.} 2010{\it a}). The results are then used for consecutive approximations of thermodynamic quantities. The procedure rapidly converges, which emphasizes the importance of the few-body aspect of many-body problems and leads to the conjecture that even at low temperatures and in the strongly interacting regime, where we lack a small parameter, the natural sorting criterion for different sets of Feynman diagrams is the number of particles (particle-hole pairs) involved.

\tableofcontents

\maintext

\chapter{The two-body problem and resonance width}

\label{Chapter: Two-body problem}

In this chapter we discuss the effective range expansion for the two-body scattering amplitude and put a special emphasis on the role of the resonance width. Near a scattering resonance the scattering length can be modified and, in particular, can take anomalously large values (i.e. $|a| \gg R_e$). Another very important parameter characterizing the resonance is its width, which is determined by the strength of the coupling between the closed and open channels. The narrower the resonance, the stronger the collision energy dependence of the scattering amplitude. In the effective range expansion this leads to an increased value of the effective range. 

In order to familiarize ourselves with the notions of scattering length, resonance width, and effective range we will consider a simple model of a single-channel potential with a barrier. It mimics the Feshbach resonance picture and clearly illustrates the role of the barrier. Then we will derive a general formula for the effective range and discuss the relation between the effective range and the probability of finding two atoms in the closed channel. This quantity plays a very important role in many problems involving Feshbach resonances with finite width. As a particular example, we will discuss the structure of a Feshbach molecule.   

\section{Potential well with delta-function barrier}
\label{sec:potwithdeltabarrier}

Consider two atoms with reduced mass $\mu$ interacting via an isotropic interaction potential $V(r)$. The radial Schr\"odinger equation for the relative motion with zero orbital angular momentum reads
\begin{equation}\label{RadSchrEq}
\left [-\frac{1}{2\mu}\frac{\partial^2}{\partial r^2}+V(r)-\frac{k^2}{2 \mu}\right ]\chi(r)=0,
\end{equation} 
where $E=k^2/2\mu$ is the collision energy and we use $\chi(r)=r\psi(r)$.

\begin{figure}[hptb]
\begin{center}
\includegraphics[width=0.6\columnwidth,clip,angle=-90]{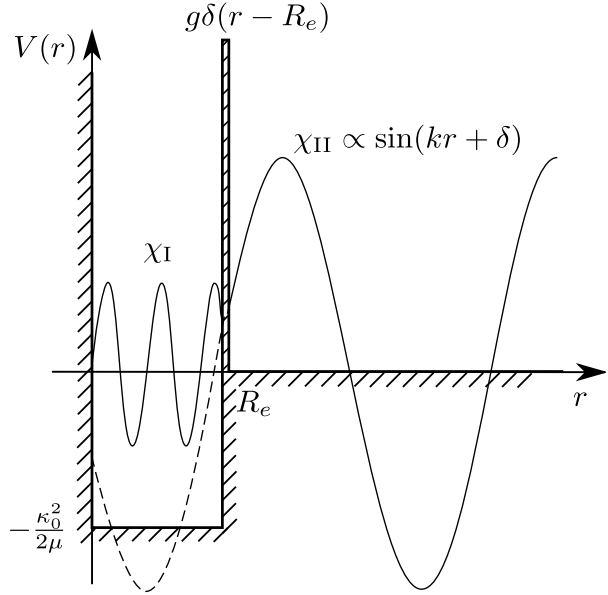}
\caption{The square well potential with a delta-function barrier (\ref{WellWithBarrier}) and the wavefunction in the inner, $\chi_{\rm I}$, and the outer, $\chi_{\rm II}$, regions at the collision energy $k^2/2\mu$ (solid line). The spike of the wavefunction at $r=R_e$ is governed by Eq.~(\ref{MatchingCond}). The dashed line shows the wavefunction $\chi_{\rm II}$ extrapolated to the inner region I.}
\label{Fig:WellWithBarrier}
\end{center}
\end{figure}

Let us take $V(r)$ in the form
\begin{equation}\label{WellWithBarrier}
V(r)=g\delta (r-R_e)-\left\{\begin{array}{ll} \kappa_0^2/2\mu,&r<R_e,\; {\rm region\; I,}\\
0,&r\ge R_e,\; {\rm region\; II.}\end{array}\right.
\end{equation} 
The potential (\ref{WellWithBarrier}) is plotted in Fig.~\ref{Fig:WellWithBarrier}. It allows one to vary the resonance width at will by varying the barrier strength $g$. The solutions of Eq.~(\ref{RadSchrEq}) in regions I and II are, respectively, $\chi_{\rm I}\propto \sin (\sqrt{\kappa_0^2+k^2}r)$ and $\chi_{\rm II}\propto \sin (kr+\delta)$. The matching condition at $r=R_e$ reads
\begin{equation}\label{MatchingCond}
\chi'_{\rm II}(R_e)/\chi_{\rm II}(R_e)-\chi'_{\rm I}(R_e)/\chi_{\rm I}(R_e)=g,
\end{equation}
which gives the phase shift $\delta$:
\begin{equation}\label{Phaseshift}
\cot \delta = \frac{\sqrt{\kappa_0^2+k^2}+\tan (\sqrt{\kappa_0^2+k^2}R_e)[g+k\tan (kR_e)]}{k-\tan (kR_e)[g+\sqrt{\kappa_0^2+k^2}\tan (\sqrt{\kappa_0^2+k^2}R_e)]}.
\end{equation}
The $s$-wave scattering amplitude is given in terms of the phase shift as (Landau and Lifshitz 1987)
\begin{equation}\label{ScatAmpl}
f(k)=1/(k\cot \delta(k)-ik).
\end{equation} 

The idea of the zero-range approximation is to extrapolate the solution $\chi_{\rm II}$ into region I (dashed line in Fig.~\ref{Fig:WellWithBarrier}) and use its logarithmic derivative at the origin to construct the zero-range boundary condition which then replaces the potential $V$:
\begin{equation}\label{ZeroRangeBC}
\chi'(0)/\chi(0)=k\cot \delta(k).
\end{equation}
By construction, the $s$-wave scattering amplitudes obtained by solving Eq.~(\ref{RadSchrEq}) on the one hand and the free-motion Schr\"odinger equation with the boundary condition (\ref{ZeroRangeBC}) on the other hand are identical. In general, substituting the potential by the boundary condition (\ref{ZeroRangeBC}) does not make the scattering problem easier since we do have to calculate the phase shift. However, the zero-range approximation becomes valuable for small momenta, when one needs only a few parameters to describe the scattering. Namely, one writes down the effective range expansion
\begin{equation}\label{EffectiveRangeExp}
k\cot \delta(k)=-1/a+(r_0/2)k^2+...,
\end{equation}
where $a$ is the scattering length, $r_0$ is the effective range, and the terms denoted by ... contain higher powers of $k$. 

Typically (i.e. for most of interatomic interactions provided by nature), the length parameters $a$ and $r_0$ are of the order of the physical range of the potential $R_e$. Then, in the ultracold limit the first term on the right hand side of Eq.~(\ref{EffectiveRangeExp}) suffices and one arrives at the so-called Bethe-Peierls boundary condition for the wavefunction
\begin{equation}\label{BP1}
\chi'(0)/\chi(0)=(r\psi(r))'/r\psi(r)|_{r=0}=-1/a,
\end{equation}
or, equivalently,
\begin{equation}\label{BP2}
\psi(r)\propto 1/r-1/a,\; r \rightarrow 0.
\end{equation}
Equation (\ref{BP2}) sets the relation between the coefficient in front of the singular term and the regular term of the wavefunction. Besides, it points to the physical meaning of the scattering length - for positive $a$ the node of the zero-energy wavefunction is found at $r=a$\footnote{Here we mean the wavefunction obtained by extrapolating its large-distance asymptote to the region $r\sim R_e$.}. In this sense, at sufficiently low energies any potential characterized by $a>0$ is equivalent to a hard-core one of radius $a$.

\begin{figure}[hptb]
\begin{center}
\includegraphics[width=0.8\columnwidth,clip,angle=0]{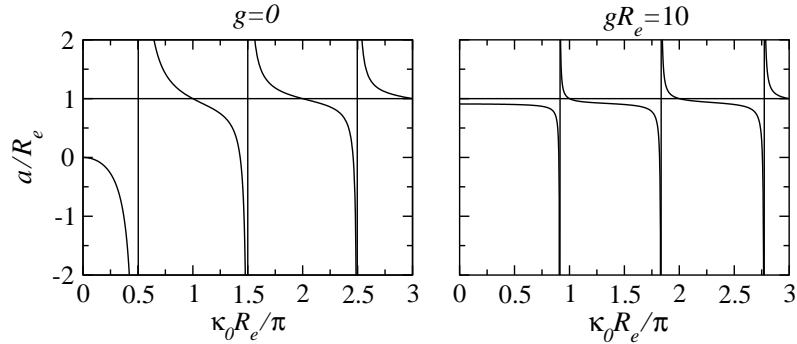}
\caption{The scattering length versus $\kappa_0$ for $g=0$ (left panel) and $gR_e=10$ (right panel).}
\label{Fig:ScatLength}
\end{center}
\end{figure}

In the case of the potential (\ref{WellWithBarrier}) the scattering length is given by
\begin{equation}\label{ScatLength}
a=R_e-1/[g+\kappa_0\cot (\kappa_0 R_e)].
\end{equation}
In Fig.~\ref{Fig:ScatLength} we show the dependence $a(\kappa_0)$ for $g=0$ (left panel) and $gR_e=10$ (right panel). We see that by changing the depth of the well we can tune the scattering length to a resonance ($a=\infty$). This happens every time a new bound state crosses the zero energy threshold. 

In the limit $g=\infty$ the regions I and II completely decouple, the former we can call closed channel, the latter -- open channel. Obviously, in this case stationary states in the closed channel live independently and the potential (\ref{WellWithBarrier}) is equivalent to a hard-core one characterized by $a= R_e$. For finite but large $g$ the stationary states at positive energies become quasi-stationary and when the collision energy coincides with the position of one of them we encounter a narrow scattering resonance. This type of resonant scattering is called the Breit-Wigner scattering or the scattering on a quasi-stationary level.

Expanding Eq.~(\ref{Phaseshift}) at small momenta one obtains the effective range $r_0$ for our model potential (\ref{WellWithBarrier}). In fact, it makes sense to discuss $r_0$ only for very large values of $a$ when $1/ak$ is comparable to $r_0k$ [see Eq.~(\ref{EffectiveRangeExp})]. In such a narrow vicinity of the resonance $r_0$ can be considered equal to its value at $a=\infty$:
\begin{equation}\label{EffectiveRange}
r_0(a=\infty)=R_e-g(1+gR_e)/\kappa_0^2 \xrightarrow[g\rightarrow \infty]{} -(g/\kappa_0)^2R_e.
\end{equation}
We see that in the limit of a narrow resonance ($g\rightarrow \infty$) the effective range $r_0$ is negative and can be much larger than the physical range of the potential. Below we will see that this statement holds generally for narrow resonances. 

Let us introduce a positive length parameter 
\begin{equation}\label{Rstar}
R^*=-r_0/2 > 0.
\end{equation}
We call a resonance narrow if $R^*\gg R_e$. Keeping only the first two terms in the effective range expansion (\ref{EffectiveRangeExp}) we write the scattering amplitude (\ref{ScatAmpl}) in the form
\begin{equation}\label{ScatAmplBW}
f(k)=-\frac{1}{1/a+R^*k^2+ik}.
\end{equation}
Equation (\ref{ScatAmplBW}) has the well-known low-energy Breit-Wigner resonance shape, which becomes apparent if we rewrite it equivalently in energy units:
\begin{equation}\label{ScatAmplBWEnergy}
f(E)=-\frac{\hbar\gamma/\sqrt{2\mu}}{E-E_{res}+i\gamma\sqrt{E}},
\end{equation}
where the position of the quasi-stationary state $E_{res}$ and the tunneling amplitude $\gamma$ are related to the lengths scales $a$ and $R^*$ as $a=-\hbar\gamma/\sqrt{2\mu}E_{res}$ and $R^*=\hbar/\sqrt{2\mu}\gamma$. The parameter $\gamma \sqrt{E}$ (multiplied by 2) is nothing else than the decay rate of the quasi-stationary state, and therefore, both $\gamma$ and $R^*$ are necessarily positive.

Similar to resonances in the model potential with barrier (\ref{WellWithBarrier}) magnetic Feshbach resonances occur when the collision energy of two atoms is close to the energy of a quasi-discrete molecular state in another hyperfine domain, in this case closed channel. The tuning of the scattering length is achieved by shifting the open and closed channels with respect to each other in an external magnetic field (hyperfine states corresponding to the open and closed channels have different magnetic moments). For magnetic resonances
\begin{equation}\label{RstarFeshbach}
R^*=\frac1{2\mu a_{\mbox{\tiny bg}}\mu_{\mbox{\tiny rel}}\Delta B},
\end{equation}
where $a_{\mbox{\tiny bg}}$ is the background scattering length, $\mu_{\mbox{\tiny rel}}$ is the difference in the magnetic moments of the closed and open channels, and $\Delta B$ is the magnetic width of the Feshbach resonance. Usually, there are many Feshbach resonances for a given atomic gas or mixture in the realistic magnetic field range and, most of the time, one has a possibility to work with a couple of ``nice'' wide resonances. However, there are systems where all available resonances are narrow. For example, parameters of the widest resonance in $^{87}$Rb (at 1007.4 G) are $\Delta_B\approx 0.17$ G, $R_e\approx 4.4$nm, $R^*\approx 32$nm (Volz {\it et al.} 2003). Similarly, all $^6$Li-$^{40}$K interspecies resonances discussed so far are characterized by $R^*\gtrsim 100$nm 
(Wille {\it et al.} 2008; Tiecke {\it et al.} 2009), whereas the van der Waals range for this alkali pair is $R_e\approx2.2$nm. The review of Chin {\it et al.}
(2010) broadly covers the phenomenon of Feshbach resonances and related physics in ultracold gases and, in particular, presents parameters for many resonances.

As we have already mentioned, the situation where $a$ is much larger than the physical range of the potential is rare and it usually requires a finetuning of the potential. Large $R^*$ arises for narrow resonances when the scattering amplitude is characterized by a strong dependence on the collision energy. One can imagine an interatomic potential for which the higher order terms in the effective range expansion (\ref{EffectiveRangeExp}) are also anomalously large. For example, we can introduce one or several additional closed channels with quasi-stationary states very close to the open channel threshold resulting in a rather exotic scattering amplitude. However, in all practically relevant cases these terms can be neglected as the typical distance between neighboring molecular states is rather large -- of order $1/\mu R_e^2$. In particular, for the model potential (\ref{WellWithBarrier}) the next term in the expansion (\ref{EffectiveRangeExp}) at $a=\infty$ and in the limit of large $g$ is proportional to $(g/\kappa_0)^4R_e^3k^4 \propto R^{*2}R_e k^4$. This means that the potential (\ref{WellWithBarrier}) has a well defined zero-range limit: tending $R_e$ to zero one can accordingly modify $\kappa_0$ and $g$ in such a way that $a$ and $R^*$ remain unchanged whereas all other terms in the effective range expansion vanish. In this limit Eqs.~(\ref{ScatAmplBW}) and (\ref{ScatAmplBWEnergy}) become exact and the Bethe-Peierls boundary condition reads
\begin{equation}\label{BP3}
\chi'(0)/\chi(0)=(r\psi(r))'/r\psi(r)|_{r=0}=-1/a-2\mu R^* E,
\end{equation}
or, in analogy with (\ref{BP2}),
\begin{equation}\label{BP4}
\psi(r)\propto 1/r-1/a-2\mu R^* E,\; r \rightarrow 0,
\end{equation}
where $E$ is the collision energy.

Clearly, $a$ is not always enough to characterize the interaction strength. For example, $a$ can be infinite, but typical energies in the problem can be so high that the interactions are off-resonant. Thus, as a more appropriate quantity it is reasonable to introduce the so-called effective energy-dependent scattering length $\tilde a(E)$ defined by the equation 
\begin{equation}\label{eq:energydependentscatlength}
1/\tilde a(E)=1/a+2\mu R^* E.
\end{equation} 
Comparison of $\tilde a(E)$ with other lengthscales in the problem gives us an idea of how strong the interaction is at a given (or typical for this problem) collision energy.

\section{Effective range and population of closed channel}

Let us now demonstrate the relation between the effective range and the probability to find two atoms within the physical range of the potential. We will then use this relation to understand the structure of a weakly bound molecular state near a narrow resonance.

Consider the Schr\"odinger equation (\ref{RadSchrEq}) with an arbitrary short-range potential $V(r)$. We require that $V(r)$ can be neglected at $r>R_e$ and that the wavefunction $\chi$ vanishes at the origin. Assume that $\chi_0$ and $\chi_k$ are two real solutions of Eq.~(\ref{RadSchrEq}) at zero and finite energy, respectively:
\begin{equation}\label{SchrZeroEn}
[-\partial^2/\partial r^2+2\mu V(r)]\chi_0(r)=0,
\end{equation} 
\begin{equation}\label{SchrFiniteEn}
[-\partial^2/\partial r^2+2\mu V(r) -k^2]\chi_k(r)=0.
\end{equation} 
We multiply Eq.~(\ref{SchrZeroEn}) by $\chi_k$ and Eq.~(\ref{SchrFiniteEn}) by $\chi_0$. Subtracting the results we obtain
\begin{equation}\label{Subtracted}
(-\chi_0\chi'_k+\chi_k\chi'_0)'=k^2\chi_0\chi_k,
\end{equation} 
Let us now introduce the functions $\tilde\chi_0$ and $\tilde\chi_k$ which satisfy, respectively, Eqs.~(\ref{SchrZeroEn}) and (\ref{SchrFiniteEn}) with $V(r)\equiv 0$ and which equal $\chi_0$ and $\chi_k$ at distances $r>R_e$. These functions satisfy
\begin{equation}\label{SubtractedTilde}
(-\tilde\chi_0\tilde\chi'_k+\tilde\chi_k\tilde\chi'_0)'=k^2\tilde\chi_0\tilde\chi_k.
\end{equation}    
We subtract Eq.~(\ref{SubtractedTilde}) from Eq.~(\ref{Subtracted}) and integrating the result from $0$ to $R_e$ we obtain
\begin{equation}\label{SubtractedSubtracted}
-\tilde\chi_0(0)\tilde\chi'_k(0)+\tilde\chi_k(0)\tilde\chi'_0(0)=k^2\int_0^{R_e} [\chi_0(r)\chi_k(r)-\tilde\chi_0(r)\tilde\chi_k(r)] dr.
\end{equation}  
In the limit $k\rightarrow 0$ we can use the boundary condition (\ref{ZeroRangeBC}) for $\tilde\chi_0$ 
and $\tilde\chi_k$ and the effective range expansion (\ref{EffectiveRangeExp}) up to the effective range term. For the effective range we obtain
\begin{equation}\label{EffRangeGeneral}
R^*=-r_0/2=\tilde\chi_0^{-2}(0)\left (\int_0^{R_e} \chi_0^2(r)dr-\int_0^{R_e}\tilde\chi_0^2(r) dr\right ).
\end{equation}
The first integral on the right hand side of Eq.~(\ref{EffRangeGeneral}) gives the probability of finding the two atoms in the region $r<R_e$:
\begin{equation}
P_{r<R_e}=4\pi \int_0^{R_e} \chi_0^2(r) dr.
\end{equation}
Near a narrow resonance $P_{r<R_e}$ is dominated by the population of the quasi-stationary state in the closed channel. The second integral in Eq.~(\ref{EffRangeGeneral}) equals
\begin{equation}
\int_0^{R_e} \tilde\chi_0^2(r) dr=\tilde\chi_0^2(0)R_e[1-R_e/a+(R_e/a)^2/3],
\end{equation}
where we have used the equality $\tilde\chi_0(r)=\tilde\chi_0(0)(1-r/a)$. In the case $R^*\gg R_e$ the second integral can be neglected and we get
\begin{equation}\label{EffRangeClosedChannelProbability}
P_{r<R_e} \approx 4\pi \tilde\chi_0^2(0) R^*.
\end{equation}

Equation (\ref{EffRangeClosedChannelProbability}) allows one to find the occupation of the closed channel (sometimes called bare molecular state) without solving coupled channel equations. One just needs to know the effective range parameter $R^*$, which can be found from Eq.~(\ref{RstarFeshbach}), and the short distance asymptote of the open channel wavefunction, $\psi(r)\approx \tilde\chi_0(0) (1/r-1/a)$. Note, that $\tilde\chi_0(0)$ is nothing else than the coefficient in front of the singular part of the wavefunction of the two-particle relative motion just outside of the potential support.

Equation (\ref{EffRangeClosedChannelProbability}) becomes exact in the zero-range limit and, together with Eqs.~(\ref{ScatAmplBW}), (\ref{ScatAmplBWEnergy}), and (\ref{BP3}) it generalizes the usual zero-range theory to the case of narrow resonances. Let us now demonstrate how this modified zero-range theory works by calculating parameters of a weakly bound molecular state near a narrow resonance. 

\section{Weakly bound molecule near a narrow resonance}
\label{MoleculeNarrowRes}

The wavefunction and the energy of a weakly bound molecular state can be found in the zero-range approximation, the applicability of which in this case requires that the size of the bound state be much larger than the physical range of the potential $R_e$. The solution of the free-motion Schr\"odinger equation
\begin{equation}\label{ShrNonInt}
[-\partial^2/\partial r^2+ \kappa^2]\chi(r)=0,
\end{equation}
decaying at large distances reads
\begin{equation}\label{MolecularWavefunction}
\chi(r)=\alpha_{\rm norm} \exp (-\kappa r),
\end{equation}
where $\alpha_{\rm norm}$ is a real normalization coefficient and $\kappa$ is determined from the Bethe-Peierls boundary condition (\ref{BP3}):
\begin{equation}\label{BP5}
\chi'(0)/\chi(0)=-\kappa=-1/a+R^* \kappa^2.
\end{equation}
This quadratic equation can also be obtained by looking for the poles of the scattering amplitude (\ref{ScatAmplBWEnergy}). One of its roots gives the energy of the bound state 
\begin{equation}\label{BoundStateEnergy}
\varepsilon_0=-\frac{\kappa^2}{2\mu}=-\frac{1}{2\mu}\left( \frac{\sqrt{1+4R^*/a}-1}{2R^*}\right)^2.
\end{equation}

The normalization constant $\alpha$ is calculated in the following way. The integral of the square of $\chi$ gives the probability of finding the atoms in the ``open'' channel $P_{\rm open}=P_{r>R_e}$ (in this case $R_e=0$):
\begin{equation}\label{ProbablityOpen}
P_{\rm open}=4\pi \int_0^\infty \chi^2(r) dr=2\pi \alpha_{\rm norm}^2 /\kappa.
\end{equation}
At the same time the closed channel population is given by Eq.~(\ref{EffRangeClosedChannelProbability}):
\begin{equation}\label{ProbablityClosed}
P_{\rm closed}=4\pi \chi^2(0) R^*=4\pi \alpha_{\rm norm}^2 R^*.
\end{equation}
Using Eq.~(\ref{BoundStateEnergy}) and the fact that $P_{\rm open}+P_{\rm closed}=1$ we get
\begin{equation}\label{ProbablityOpenResult}
P_{\rm open}=1/\sqrt{1+4R^*/a},
\end{equation}
and
\begin{equation}\label{NormalizationAlpha}
\alpha_{\rm norm}=\sqrt{\frac{\kappa}{2\pi (1+2R^*\kappa)}}=\sqrt{\frac{1-1/\sqrt{1+4R^*/a}}{4\pi R^*}}.
\end{equation}

Depending on the value of the ratio $R^*/a$ we distinguish two regimes: the case $R^*/a\ll 1$ we call the regime of small detuning, and the opposite limit -- the regime of intermediate detuning. To be more precise, in the latter case we require $R_e\ll a\ll R^*$ since we would like to stay within the region of validity of the zero-range approximation. In the regime of small detuning the molecular energy equals $\varepsilon_0\approx-1/2\mu a^2$ and the probability of finding the atoms in the closed channel is small. The physics in this case is similar to the one near a wide resonance. In the regime of intermediate detuning $\varepsilon_0\approx-1/2\mu R^*a\approx E_{res}$ and $P_{\rm closed}\approx 1$, which means that the molecular state in this case is dominated by the closed channel or bare molecular state.  

\chapter{Basics of the three-body problem with short-range interactions}

\label{Chapter: Three-body problem}

In this chapter we apply the zero-range approach to the problem of three atoms. We consider the phenomenon of the three-body collapse, or Thomas collapse, and define the Efimov effect. Then we introduce a model system of one light and two heavy atoms where one can use the Born-Oppenheimer approximation. The Efimov physics in this model is extremely transparent and one can understand the role of the two-body interaction parameters (scattering length, resonance width), masses of the atoms, and their quantum statistics. We also discuss the question of inelastic three-body losses.

\section{Thomas collapse}

Consider three identical bosons interacting with each other via the square well potential (\ref{WellWithBarrier}) with $g=0$. Let us gradually decrease $R_e$ and at the same time increase the depth of the potential in such a way that the scattering length $a$ is unchanged. In this way we arrive at a zero-range potential equivalent to the initial one in the ultracold limit. What happens with the spectrum of the three-body problem as we decrease $R_e$? Below we show that its ground state energy is unbound from below in the limit $R_e\rightarrow 0$. This phenomenon was discovered by Thomas (1935) who theoretically addressed the mass defect of tritium and arrived at a number of important conclusions on the character of the proton-neutron and neutron-neutron interaction.

The Hamiltonian for three bosons in the center-of-mass reference frame reads (we set $\hbar=m=1$)
\begin{equation}\label{HamEq3Bosons}
H=-\nabla_{\bf x}^2-\nabla_{\bf y}^2 + V(y)+\sum_\pm V(|\sqrt{3}{\bf x}\pm {\bf y}|/2),
\end{equation}
where ${\bf y}$ is the distance between two of them and $\sqrt{3}{\bf x}/2$ is the distance from their center of mass to the third boson. Consider the variational wavefunction
\begin{equation}\label{Gaussian}
\Psi_{\rm var}({\bf x,y})=(\omega/2\pi)^{3/2}\exp [-\omega(x^2+y^2)/4],
\end{equation}
which is the normalized ground state wavefunction of a six-dimensional harmonic oscillator of mass 1/2 and frequency $\omega$ -- our variational parameter. The ground state energy of the Hamiltonian (\ref{HamEq3Bosons}) satisfies
\begin{eqnarray}\label{VariationalInequality}
E_0 & \leq & \langle \Psi_{\rm var}|H|\Psi_{\rm var}\rangle=3\omega/2+3(\omega/2\pi)^{3/2} \int V(y) \exp (-\omega y^2/2){\rm d}^3 y \nonumber \\
 & = & 3\omega/2-3\kappa_0^2 [{\rm erf} \sqrt{\omega R_e^2/2}-\sqrt{2\omega R_e^2/\pi}\exp (-\omega R_e^2/2) ],
\end{eqnarray} 
where we use Eq.~(\ref{WellWithBarrier}) and the fact that the three interaction terms in (\ref{HamEq3Bosons}) contribute equally to the matrix element $\langle \Psi_{\rm var}|H|\Psi_{\rm var}\rangle$.

When we reach the condition $R_e \ll |a|$ the depth of the potential becomes practically independent of the exact value of $a$ and is determined by the equation $\cot (\kappa_0 R_e)=0$ [see Eq.~(\ref{ScatLength})]. The minimum depth of the potential is then given by $\kappa_0=\pi/2R_e$. Substituting this value to Eq.~(\ref{VariationalInequality}) we find that its right hand side is minimized for $\omega\approx 2.1/R_e^2$ and we get
\begin{equation}\label{VariationalInequalityFinal}
E_0\leq -0.167/R_e^2,
\end{equation} 
which confirms the statement that the ground state energy is unbound from below as $R_e\rightarrow 0$.

In the case of three bosons this result is expected: The inequality $|a|\gg R_e$ means that there is a bound or quasi-bound two-body state. In 3D, this requires that the potential is of a certain finite attractive strength. If we {\it a priori} assume that two of the atoms are very close to each other, the third atom will experience twice the attractive potential and will then be tightly bound, which verifies the initial assumption. Interestingly, Thomas showed that the statement holds for the proton-neutron-neutron (tritium) system in which the neutron-neutron interaction is neglected. In this case the above hand waving argument does not work because the interaction potential in between a neutron and a deutron is the same as the neutron-proton interaction. However, the Thomas effect can be explained by the larger effective mass of the neutron-deutron system compared to the neutron-proton one. The strict variational analysis is rather bulky in this case (Thomas 1935).   

The system of three ${}^4$He atoms is one of the most spectacular examples in which this physics actually takes place. The ${}^4$He-${}^4$He potential is just attractive enough to accommodate a single bound state close to the continuum, the scattering length is positive and is an order of magnitude larger than the range of the potential. However, the ${}^4$He${}_3$ trimer is much smaller in size and significantly stronger bound than the dimer. In fact, even if the interatomic potential were a bit weaker, so that the dimer state is just pushed into the continuum, the trimer state would still be there. This phenomenon is called the Borromean binding -- none of the two-body subsystems has a bound state, but there exists a trimer state. An example of the Borromean binding is the ${}^6$He nuclei which is stable in spite of the fact that neither $\alpha$-$n$ nor $n$-$n$ system is bound.

We now know that the ground state of such systems is a trimer state with the binding energy $\sim 1/R_e^2$. Are there other trimer states and what does their spectrum look like?

\section{Efimov effect and discrete scaling invariance}

Consider again three identical bosons interacting with each other via a potential of the range $R_e$. This time let us keep $R_e$ constant, but increase the value of $a$, no matter positive or negative, by slightly tuning the depth of the potential. Efimov showed (Efimov 1970) that as $a\rightarrow \infty$ the number of trimer states is proportional to $(s_0/\pi)\log (|a|/R_e)\rightarrow \infty$, where $s_0$ is a dimensionless parameter. A new trimer state emerges from the continuum each time $a$ is increased by a factor of $\exp(\pi/s_0)$. When $a=\infty$ we have infinite number of trimer states with binding energies $|\varepsilon_n|\propto \exp (-2\pi n/s_0)$.

The effect is quite general -- it can take place in a system of three particles with different masses and statistics as long as at least two of the three interactions are resonant. The parameter $s_0$ depends on the masses and statistics of the atoms, for identical bosons it equals $s_0 \approx 1.00624$ so that the scaling parameter $\exp(\pi/s_0)\approx 22.7$. 

In fact, the Efimov effect goes along with the phenomenon of discrete scaling invariance: a three-body observable of the dimension of length to the power $\alpha$ close to the resonance ($a \rightarrow \infty$) behaves as $a^\alpha f(a/R_e)$, where $f$ is a dimensionless function periodic on the logarithmic scale, i.e. $f(x)= f(x\exp[\pi/s_0])$. 

We are used to the idea of continuous scaling invariance: once we multiply all lengthscales in the Hamiltonian by a factor of $\lambda$, all eigenenergies get multiplied by $1/\lambda^2$ and all eigenfunctions trivially rescale to satisfy the normalization condition. It is important to keep in mind that in the example discussed above we do change $a$, but we do not rescale $R_e$. Therefore, the discrete scaling invariance does not contradict our understanding of the continuous scaling. As we show below, when $R_e\rightarrow 0$ the short-range physics does not completely drop out of the (three-body) problem.

\section{Born-Oppenheimer approximation}

Let us consider a light atom of mass $m$ interacting resonantly via a short-range potential with two heavy atoms of mass $M$. For $M/m\gg 1$ one can use the Born-Oppenheimer approximation to illustrate the essential physics leading to the Efimov effect (Fonseca {\it et al.} 1979). The Born-Oppenheimer approach takes advantage of the large mass ratio by assuming that the state of the light atom adiabatically adjusts itself to the distance ${\bf R}$ between the heavy atoms. The three-body problem thus splits into two ``one-body'' pieces. First, we solve the problem of the light atom in the field of two fixed scatterers at positions ${\bf R}/2$ and $-{\bf R}/2$ (a version of the double-well potential problem). The binding energy of the light atom then serves as the interaction potential between the heavy ones. The second step is then to solve the Schr\"odinger equation for the relative motion of the heavy atoms.

If there were only one heavy atom at position ${\bf R}/2$, the wavefunction of the light atom could be written as
\begin{equation}\label{eq:BOwavefunctionOneHeavy}
\psi({\bf r}) \propto \frac{e^{-\kappa|{\bf r-R}/2|}}{|{\bf r-R}/2|},
\end{equation}
where $\kappa$ satisfies Eq.~(\ref{BP5}). In the case of two heavy atoms we look for the solution in the form
\begin{equation}\label{eq:BOwavefunction}
\psi_{\bf R}({\bf r}) \propto C_1 \frac{e^{-\kappa(R)|{\bf r-R}/2|}}{|{\bf r-R}/2|}+C_2 \frac{e^{-\kappa(R)|{\bf r+R}/2|}}{|{\bf r+R}/2|}.
\end{equation}
The wavefunction (\ref{eq:BOwavefunction}) satisfies the Schr\"odinger equation
\begin{equation}\label{BOSchrodinger}
\left[-\frac{1}{2m}\nabla_{\bf r}^2+\sum_\pm V(|{\bf r \pm R}/2|)\right]\psi_{\bf R}({\bf r})=\epsilon(R)\psi_{\bf R}({\bf r})
\end{equation}
with the energy 
\begin{equation}\label{eq:BOenergy}
\epsilon(R)=-\kappa^2(R)/2m.
\end{equation}
The parameters $\kappa(R)$, $C_1$, and $C_2$ are obtained by applying the Bethe-Peierls boundary condition (\ref{BP3}) to $\psi_{\bf R}({\bf r})$ at vanishing ${\bf r \pm R}/2$ and by using the fact that the collisional energy equals $\epsilon(R)$. In particular, for the boundary ${\bf r - R}/2\rightarrow 0$ the resulting equation reads
\begin{equation}\label{EquationForC}
[\kappa(R)-1/a+R^*\kappa^2(R)]C_1-\{\exp [-\kappa (R)R]/R]\}C_2=0,
\end{equation}
and for the other boundary we get the same equation with interchanged $C_1$ and $C_2$. 

\begin{figure}[hptb]
\begin{center}
\includegraphics[width=0.6\columnwidth,clip]{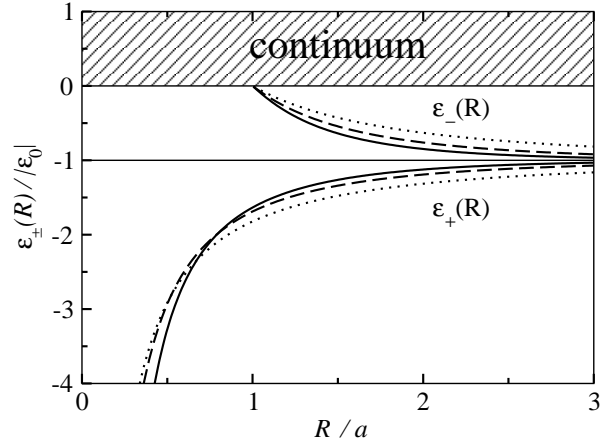}
\caption{Born-Oppenheimer potentials $\epsilon_-$ and $\epsilon_+$ in units of the molecular binding energy $|\epsilon_0|$ for different detunings: $R^*/a=0$ (solid), $R^*/a=1$ (dashed), and $R^*/a=10$ (dotted).}
\label{fig:BOPotentials}
\end{center}
\end{figure}

These equations have two solutions: $\{C_1,C_2\}_+=\{1,1\}$ and $\{C_1,C_2\}_-=\{1,-1\}$. Substituting them into Eq.~(\ref{eq:BOwavefunction}) we obtain the wavefunctions $\psi_{{\bf R},+}({\bf r})$ and $\psi_{{\bf R},-}({\bf r})$, which are nothing else than the symmetric and antisymmetric bound states of our double-well potential problem. The equation for $\kappa_\pm (R)$ reads
\begin{equation}\label{eq:kappa}
\kappa_\pm(R)\mp \exp\left[-\kappa_\pm(R)R\right]/R= 1/a-R^*\kappa_\pm^2(R).
\end{equation}
Solving Eq.~(\ref{eq:kappa}) and using Eq.~(\ref{eq:BOenergy}) we get the induced interaction potentials $\epsilon_\pm(R)$ for the heavy atoms. In Fig.~\ref{fig:BOPotentials} we plot these potentials for different values of $R^*/a$. We observe that $\epsilon_+(R)$ is purely attractive and $\epsilon_-(R)$ is repulsive. In the limit of large $R$ both potentials tend to the molecular binding energy (note that the Born-Oppenheimer value of $\epsilon_0$ differs from Eq.~(\ref{BoundStateEnergy}) by the factor $\mu /m$). At distances $R<a$ the symmetric state (+) is the only bound state since the antisymmetric one (-) goes to the continuum at $R=a$.

The second step of the Born-Oppenheimer method consists of solving the Schr\"odinger equation for the heavy atoms by using $\epsilon_\pm(R)$ as the potential energy surface. Let us postpone the question on how this induced interaction depends on $R^*/a$ and first consider the case $R^*=0$. Then, in the limit $R\gg a$ the parameter $\kappa_\pm(R)$ can be expanded in a power series in the small parameter $\exp(-R/a)$. The first three terms read
\begin{equation}\label{eq:kappaLargeRWideRes}
\kappa_\pm(R)\approx a^{-1}\pm R^{-1}\exp(-R/a) - R^{-1}\exp(-2R/a),
\end{equation}
which gives the long-distance asymptotes:
\begin{equation}\label{eq:epsilonLargeRWideRes}
\epsilon_\pm(R)\approx -\frac{1}{2ma^2}\mp \frac{\exp(-R/a)}{maR}+\frac{\exp(-2R/a)}{maR}(1-a/2R).
\end{equation}
This means that at large distances the exchange of the light atom leads to a Yukawa-type force between the heavy atoms. In the limit $R\ll a$ only the symmetric state is bound, the right hand side of Eq.~(\ref{eq:kappa}) can be neglected, and we obtain $\kappa_+(R)\approx C/R^2$, where $C\approx 0.567$ is the solution of the transcendental equation $C=\exp(-C)$. Accordingly, the short-distance asymptote of $\epsilon_+(R)$ reads
\begin{equation}\label{eq:epsilonShortRWideRes}
\epsilon_+(R)\approx -\frac{C^2}{2mR^2}\approx -\frac{0.16}{mR^2}.
\end{equation}
The $1/R^2$-potential does not have a characteristic length-scale and is at the origin of the Efimov effect. Note that in the case $a=\infty$ Eq.~(\ref{eq:epsilonShortRWideRes}) holds for all $R$.

\section{Efimov effect in heavy-heavy-light system}
\label{EfimovBO}
Let us assume that the light atom is in the symmetric state and that it follows adiabatically the displacement of the heavy atoms ($M/m\gg 1$). Then, we can integrate out its motion and write down the Schr\"odinger equation for the heavy atoms in the form
\begin{equation}\label{BOSchrEq}
[-\nabla^2_{\bf R}/M+\epsilon_+(R)-E]\phi({\bf R})=0.
\end{equation}
At small $R$ we can neglect $E$ compared to other terms in Eq.~(\ref{BOSchrEq}) and use the short-distance asymptote of the potential (\ref{eq:epsilonShortRWideRes}). The radial Schr\"odinger equation then reads 
\begin{equation}\label{BOSchrEqShortDist}
(-\partial^2/\partial R^2 +\beta/R^2)\chi(R)=0,
\end{equation}
where $\beta=l(l+1)-0.16 M/m$ and $\chi(R)$ is the radial part of the $l$-th angular momentum component of $R\phi({\bf R})$.

Equation~(\ref{BOSchrEqShortDist}) is satisfied by the function $R^\nu$, where the power $\nu$ can take two values
\begin{equation}\label{BOnu}
\nu_{1,2}=1/2\pm\sqrt{\beta+1/4}.
\end{equation}
The case $\beta<-1/4$ corresponds to the well-known problem of the fall of a particle to the center in an attractive $R^{-2}$ potential (Landau and Lifshitz 1987). In this case a real solution of Eq.~(\ref{BOSchrEqShortDist}) is a linear superposition of $R^{\nu_1}$ and $R^{\nu_2}$ with complex conjugate coefficients. It can be expressed in the form
\begin{equation}\label{BOCosLog}
\chi(R)\propto\sqrt{R}\cos[s_0\log (R/R_0)],
\end{equation}
where $s_0=\sqrt{-\beta-1/4}$ and $R_0$ is the three-body parameter, which fixes the phase in between the oscillations of $R^{\nu_1}$ and $R^{\nu_2}$. 

The three-body parameter is sensitive to the short-range (high-energy) details of the interatomic potentials, i.e. two systems with different interparticle potentials are, generally, characterized by different three-body parameters even if the corresponding scattering lengths are the same. The task of calculating $R_0$ for a realistic potential is very challenging, but, most of the time, we need to know only a few general properties of this parameter. For example, for one and the same system it is reasonable to assume that the three-body phase is a continuous function of the potential depth, and in a sufficiently narrow vicinity of a two-body resonance it can be approximated by a constant. It then enters the zero-range theory as a fixed external parameter, whereas the scattering length changes in a very wide range. 

In the case of finite $R_e$ and $a$ the wavefunction (\ref{BOCosLog}) remains a good approximation for distances $R_e\lesssim R \lesssim |a|$. We see that it has approximately $N_b\approx (s_0/\pi)\log |a|/R_e$ nodes, which is an estimate of the number of bound states (trimer states). This number becomes infinite in the limit $a\rightarrow \infty$ exactly as predicted by Efimov.

Let us now look at the spectrum of trimers in the case $a=\infty$. Equation (\ref{BOSchrEq}) in this case reads
\begin{equation}\label{BOSchrEqInfiniteA}
(-\partial^2/\partial R^2 +\beta/R^2-ME)\chi(R)=0.
\end{equation}
Let us assume that there is a bound state with energy $E=-E_0<0$. The wavefunction $\chi_0(R)$ corresponding to this state is a linear superposition of $\sqrt{R}J_{is_0}(i\sqrt{ME_0}R)$ and its complex conjugate, where $J$ is the Bessel function. The dependence of $E_0$ on the three-body parameter is established by requiring that $\chi_0(R)$ be finite for large $R\gg 1/\sqrt{ME_0}$ and take the form (\ref{BOCosLog}) in the opposite limit. In principle, $E_0$ itself can be used as a three-body parameter. 

Let us now consider a state given by the wavefunction $\chi_0(\lambda R)$. It obviously satisfies Eq.~(\ref{BOSchrEqInfiniteA}) with energy $E=-\lambda^2E_0$ and decays at large $R$. The parameter $\lambda$ is not arbitrary -- Eq.~(\ref{BOCosLog}) gives the condition $s_0\log\lambda =\pi n$, where $n$ is an integer. This is one of the consequences of the discrete scaling invariance. We understand now that all the Efimov trimers are self-similar and their spectrum is geometric:
\begin{equation}\label{BOGeomSpectr} 
E_n=E_0\exp(2\pi n/s_0).
\end{equation}

Another manifestation of the discrete scaling invariance is the log-periodic dependence of three-body observables on the scattering length. This can be shown by rescaling the coordinates $R=a \tilde R$. In the new coordinates we have to solve the three-body problem for the unit scattering length and use the small-$\tilde R$ three-body condition
\begin{equation}\label{BOCosLogTilde}
\chi(\tilde R)\propto\sqrt{\tilde R}\cos[s_0\log (\tilde R/\tilde R_0)],
\end{equation}
where the rescaled three-body parameter depends on the scattering length, $\tilde R_0=R_0/a$. We then see that all the three-body observables should be log-periodic functions of $a$ (besides the ``trivial'' dimensional scaling with $a$) with the discrete scaling factor $\exp(\pi/s_0)$.

As an example let us consider the $s$-wave atom-molecule scattering. In order to calculate the corresponding scattering length we take the $l=0$ component of Eq.~(\ref{BOSchrEq}) with $E=-|\epsilon_0|=-1/ma^2$, i.e. with the atom-molecule collision energy equal to zero. In the rescaled coordinates, $\tilde R = R/a$, this equation reads
\begin{equation}\label{BOSchrEqInfiniteARescaled}
(-\partial^2/\partial \tilde R^2 +M\epsilon_{+,a=1}(\tilde R)+M/m)\chi(\tilde R)=0.
\end{equation}
Let us take the solution of Eq.~(\ref{BOSchrEqInfiniteARescaled}) that at $\tilde R \ll 1$ behaves as $\tilde R^{1/2+is_0}$. Its long-range asymptote is a certain linear superposition of free solutions $\tilde R^0$ and $\tilde R^1$. We write it as $A+B\tilde R$, where $A$ and $B$ are complex coefficients, which could be in principle found numerically. Since Eq.~(\ref{BOSchrEqInfiniteARescaled}) is real, the second linearly independent solution can be constructed as the complex conjugate of the first one. Then we should take their linear combination satisfying Eq.~(\ref{BOCosLogTilde}) and compare its long-range asymptote with the expression $\propto (\tilde R-\tilde a_{\rm ad})$, where $\tilde a_{\rm ad}$ is the rescaled atom-dimer scattering length. Performing this operation we find that 
\begin{equation}\label{BOAtomDimerScatLength}
\tilde a_{\rm ad}(\tilde R_0)=-\frac{A \tilde R_0^{-is_0}+A^*\tilde R_0^{is_0}}{B \tilde R_0^{-is_0}+B^* \tilde R_0^{is_0}}=\xi+\zeta \tan (s_0\log \tilde R_0 + \delta),
\end{equation}
where $\xi$, $\zeta$, and $\delta$ are real numbers. We see that the nonrescaled atom-molecule scattering length has the form $a_{\rm ad}=a\tilde a_{\rm ad}(R_0/a)$, i.e. it is indeed a log-periodic function of $a$ (up to the dimensional prefactor). The resonances in the atom-dimer scattering length denote the emergence of a new trimer state each time the scattering length is multiplied by $\exp(\pi/s_0)$.

We emphasize once again that the discrete scaling invariance is a consequence of the fact that we assume $R_0$ to be constant while we modify $a$. This is a good approximation close to a Feshbach resonance, however, theoretically we could rescale both the depth and the range of the potential so that $R_0$ is proportional to $a$. Then, we would observe the usual continuous scaling, i.e. the self-similarity for any value of the scaling factor.

\section{Role of quantum statistics and masses of atoms}
\label{sec:QuantumStatistics}

In the Born-Oppenheimer approach we can also understand the role played by the quantum statistics of the heavy atoms. Indeed, the total three-body wavefunction is proportional to the product $\phi({\bf R})\psi_{{\bf R},\pm}({\bf r})$. It should be symmetric (antisymmetric) with respect to the permutation ${\bf R}\rightleftarrows -{\bf R}$ if the heavy particles are identical bosons (fermions). Then the symmetry of $\phi$ depends
on the choice of the light atom wavefunction. Remember that $\psi_{{\bf R},+}({\bf r})$ is symmetric and $\psi_{{\bf R},-}({\bf r})$ is antisymmetric with respect to the permutation of the heavy particles. Since only the symmetric state leads to the induced attraction, the Efimov effect is expected for even angular momentum channels in the case of bosonic heavy atoms and for odd ones in the case of fermions. Of course, if the atoms are distinguishable, even if they are different hyperfine components of one and the same isotope, all angular momenta are allowed.

Although the Born-Oppenheimer approximation is supposed to work only for large mass ratios, exact calculations show that it captures quite well the essentials of the Efimov physics even for moderate $M/m$. In particular, the formula (see Section~\ref{EfimovBO})
\begin{equation}\label{eq:s0}
s_0=\sqrt{0.16 M/m-(l+1/2)^2}
\end{equation}
correctly predicts that the scaling factor $\lambda=\exp(\pi/s_0)$ decreases with the mass ratio and increases with the angular momentum. The explanation of this phenomenon in the Born-Oppenheimer picture is straightforward: an increase in the mass ratio strengthens the exchange $1/R^2$-attraction compared to the kinetic energy of the heavy atoms. In turn, the centrifugal barrier competes with the exchange attraction and prevents the heavy particles from falling to the center.

\begin{figure}[hptb]
\begin{center}
\includegraphics[width=0.6\columnwidth,clip,angle=0]{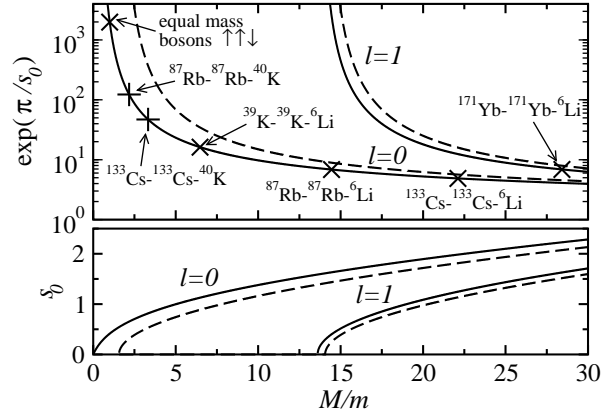}
\caption{Scaling factor $\exp(\pi /s_0)$ (upper panel) and $s_0$ (lower panel) versus $M/m$. The dashed lines are calculated by using the Born-Oppenheimer approximation and the solid lines are results of the exact approach of Sec.~\ref{Sec:Exacts0}. Symbols show the scaling factors for several realistic three-atom systems.}
\label{Fig:s0}
\end{center}
\end{figure}

In Fig.~\ref{Fig:s0} we plot the scaling factor $\exp(\pi /s_0)$ and $s_0$ versus $M/m$ for $l=0$ and $l=1$. The result of Eq.~(\ref{eq:s0}) (dashed lines) is compared to the exact one (solid lines), the derivation of which will be discussed in Sec.~\ref{Sec:Exacts0}. In the case of heavy bosons ($l=0$) the Efimov effect always takes place. However, large mass ratios are more favorable for its observation as the scaling factor $\lambda$ rapidly decreases with $M/m$, which then allows for a larger number of log-periods within a fixed window of $a$.

In the case of heavy fermions ($l=1$) we see that the Efimov effect does not happen for $M/m<13.6$ because the centrifugal barrier for the heavy fermions is stronger than the attraction induced by the light atom. The absence of the Efimov effect has advantages: First of all, the three-body problem in {\it non-Efimovian} cases is quite interesting on its own. For example, there can be trimer states which are qualitatively different from the Efimov trimers (Kartavtsev and Malykh 2007). Another striking feature of non-Efimovian systems is their collisional stability -- the reason why a two-component strongly interacting fermionic mixture is long-lived even at unitarity, $a=\infty$, whereas bosonic systems decay very rapidly near a Feshbach resonance. Let us now discuss the three-body recombination to deeply bound states -- one of the most important loss mechanisms in a strongly interacting gas.

\section{Three-body recombination into deeply bound states: Efimovian case}
\label{sec:RecombinationBosons}

In the absence or in the case of a strong suppression of two-body inelastic channels a third body is needed for a pair of atoms to form a bound state. The binding energy is then transfered into the kinetic energy of recombination products (an atom and a molecule). If the final bound state is deeply bound (typically of size $R_e$), the products are not ultracold and they escape from the trapped sample. The frequency of three-body recombination events in a unit volume is proportional to the number of all possible triples in this volume. For example, for a one-component Bose gas it is proportional to $n^3$, where $n$ is the density. The three-body recombination rate $\nu$ is the frequency per atom, it is thus proportional to $n^2$. The coefficient $\alpha_{\rm rec}$ in the equation $\nu=\alpha_{\rm rec} n^2$ is called the three-body recombination rate constant and has the dimension cm$^6$s$^{-1}$. The loss rate equals $3\nu$ as three atoms are lost at each event, and the density evolves according to the equation $\dot{n} = - 3\nu n^3$. 

The recombination to deeply bound states can also happen in collisions of atoms and weakly bound molecules. In this case it is called collisional relaxation. Its frequency per unit volume is proportional to the product of the atomic and molecular densities, $n_a n_m$. Thus, the relaxation rate per atom equals $\alpha_{\rm rel}n_m$, and per molecule, $\alpha_{\rm rel}n_a$. The coefficient $\alpha_{\rm rel}$ has dimensions cm$^3$s$^{-1}$. 

Clearly, the task of calculating or, at least, estimating $\alpha_{\rm rec}$ and $\alpha_{\rm rel}$ is very important from the practical viewpoint. The knowledge of these constants allows one to tune experimental parameters and maximize the gas longevity. On the other hand, these parameters contain a lot of information on few-body correlations in the gas. In particular, the Efimov effect has so far been observed exclusively by measuring losses as a function of the scattering length.

We now discuss the zero-range approach to this type of inelastic processes that happen at very short distances $\sim R_e$. As an illustrative example let us consider the relaxation in atom-dimer collisions and calculate the corresponding rate constant $\alpha_{\rm rel}$. It turns out that in the Efimovian case this problem can be solved rather elegantly by allowing the three-body parameter to be complex (Braaten and Hammer 2007). Indeed, the solution of the heavy-atom Schr\"odinger equation at distances $R_e\ll R \ll a$ is the sum of the incoming wave, $\sqrt{R}(R/R_0)^{-is_0}$, and the outgoing one, $\sqrt{R}(R/R_0)^{is_0}$. In the presence of the relaxation to deeply bound states the outgoing flux should be a certain fraction of the incoming one. Theoretically this is modelled by introducing the so-called elasticity parameter $\eta_*$ and writing  $R_0$ in the form $R_0=|R_0|\exp(-i\eta_*/s_0)$. The modulus of $R_0$ then fixes the relative phase of the incoming and outgoing waves, and $\eta_*>0$ ensures that the incoming flux is by the factor $\exp{4\eta_*}$ larger than the outgoing one. Everything that we have said about the three-body parameter holds also for the elasticity parameter. Namely, it is very difficult to calculate $\eta_*$ {\it ab initio}, but in a sufficiently narrow vicinity of the resonance we can approximate it by a constant, which is inserted into the zero-range theory as a parameter.

Now, let us consider a weakly bound molecule and an atom in a unit volume. The interaction energy shift for this system equals $\Delta E=2\pi a_{\rm ad}/\tilde\mu$, where $\tilde\mu\approx M/2$ is the atom-molecule reduced mass and $a_{\rm ad}$ is the atom-dimer scattering length given by Eq.~(\ref{BOAtomDimerScatLength}). Once we introduce a finite elasticity parameter [$\tilde R_0=|\tilde R_0|\exp(-i\eta_*/s_0)$ in Eq.~(\ref{BOAtomDimerScatLength})] the atom-dimer scattering length acquires an imaginary part, which in turn leads to the shift of $\Delta E$ into the lower complex half-plane. We thus see that a single atom-dimer pair decays with the rate $-2{\rm Im} (\Delta E)$, which in this case equals the relaxation rate constant
\begin{equation}\label{BORelRateConstant}
\alpha_{\rm rel,bosons}=-\frac{4\pi{\rm Im}a_{\rm ad}}{\tilde\mu}=\frac{4\pi a}{\tilde\mu}\zeta\frac{\sinh\eta_*\cosh\eta_*}{\cos^2(s_0\log|R_0/a|+\delta)+\sinh^2\eta_*},
\end{equation}
where $\zeta$ and $\delta$ are the same as in Eq.~(\ref{BOAtomDimerScatLength}). We see that $\alpha_{\rm rel,bosons}/a$ is a log-periodic function of $a$. Quite naturally it reaches its maximum at the atom-dimer scattering resonances when the cosine in the denominator in the right hand side of Eq.~(\ref{BORelRateConstant}) vanishes.

In general, whenever we know the dependence of an arbitrary three-body observable on the real three-body parameter, we can analytically continue this function in the complex $R_0$-plane thus obtaining the result for the situation when the relaxation is present. As an exercise we propose to calculate the energies and lifetimes of the Efimov states in the case $a=\infty$.

The case $\eta_*=\infty$ is quite peculiar. It corresponds to the complete absorption -- only the incoming wave is present at short distances in the three-body wavefunction. In this case we recover the continuous scaling invariance, but three-body observables acquire a significant imaginary part. For example, the atom-dimer scattering length is complex and is simply proportional to $a$ without log-periodicity.  

The real and imaginary parts of $R_0$ can be found by fitting experimental loss data. Existing experimental measurements for various isotopes and Feshbach resonances are consistent with values of $\eta^*$ in between 0.1 and 0.4. 

\section{Three-body recombination into deeply bound states: non-Efimovian case}
\label{sec:RecombinationFermions}

The non-Efimovian case corresponds to $\beta>-1/4$ in Eq.~(\ref{BOnu}). Then the three-body wave function at distances $R_e\ll R \ll a$ is the sum 
\begin{equation}\label{BOPsiNonEfimovian}
\chi(R)\propto A_1 (R/R_e)^{\nu_1}+A_2 (R/R_e)^{\nu_2},
\end{equation}
where $\nu_1>\nu_2$. Expression (\ref{BOPsiNonEfimovian}) should be matched with the solution of the three-body problem at $R\lesssim R_e$. To find the latter is challenging. However, in the absence of a three-body resonance the matching procedure implies that both terms on the right hand side of Eq.~(\ref{BOPsiNonEfimovian}) are of the same order of magnitude, i.e. $A_1\sim A_2$. Then, at distances $R\gg R_e$ the second term can be neglected compared to the first one. Note that we use exactly the same argument when we neglect the interaction in between identical fermions in the ultracold limit: out of the two possible free solutions of the $p$-wave radial Schr\"odinger equation, $R^1$ and $R^{-2}$, we choose the former one (the one which increases with $R$).

The zero-range approach to calculating the relaxation into deeply bound states in non-Efimovian cases is perturbative. It uses the unperturbed three-body wavefunction to predict the probability of finding three atoms at distances $\sim R_e$ and gives the functional dependence of $\alpha_{\rm rel}$ on the scattering length. This means that if the relaxation rate constant is known (measured) for a certain $a$, one can predict its value for any other $a\gg R_e$. 

As a specific example let us estimate the atom-dimer relaxation rate in $p$-wave collisions of heavy fermions with heavy-light molecules in the Born-Oppenheimer approximation. In this case the light atom is in the state $\psi_{{\bf R},+}({\bf r})$, which means that at $R\ll a$ it is always close to the heavy fermions. Then we just need to calculate the probability to find the heavy atoms at distances $R\sim R_e$. In order to do this we take the short-distance ($R\lesssim a$) asymptote of the heavy-atom wavefunction, $\Phi({\bf R})\approx C R^{\nu_1-1}\cos\theta_{\bf k,R}$, and match it with the long-distance one, $\Phi({\bf R})\approx \sqrt{2}\sin({\bf k}{\bf R})$. The latter is properly normalized to a single atom-molecule pair in a unit volume (we assume that we are far from any atom-molecule $p$-wave resonance), ${\bf k}$ is the relative atom-dimer momentum, and $\theta_{\bf k,R}$ is the angle between ${\bf k}$ and ${\bf R}$. Matching the short- and long-distance asymptotes at $R\sim a$ we obtain $C\sim ka^{2-\nu_1}$, which gives the probability $P_{R\lesssim R_e}\sim (R_e/a)^{2\nu_1-2}(ka)^2R_e^3$. In order to get a dimensional estimate of $\alpha_{\rm rel,fermions}$ we can multiply $P_{R\lesssim R_e}$ by the frequency of recombination processes that would take place if all three atoms are confined to distances $\sim R_e$. This frequency is of order $1/mR_e^2$. We thus obtain 
\begin{equation}\label{BORRCFermions}
\alpha_{\rm rel,fermions}=(R_e/m)(R_e/a)^{2\nu_1-2}(ka)^2.
\end{equation}
The three factors in Eq.~(\ref{BORRCFermions}) are interpreted as follows: The factor $R_e/m$ is of the order of the relaxation rate constant for $s$-wave collisions of atoms and deeply bound molecules (with the size $\sim R_e$). The factor $(R_e/a)^{2\nu_1-2}$ is the suppression factor that comes from the fact that the atoms have to tunnel under the effective repulsive potential, which is the sum of the centrifugal barrier and the exchange attraction. The power $\nu_1$ depends on the mass ratio. The factor $(pa)^2$ corresponds to the low-energy Wigner law for reactions with unit angular momentum ($l=1$ in this case). 

Accordingly, the relaxation rate constant for $s$-wave atom-molecule collisions has the form (\ref{BORRCFermions}) without the last factor. Unfortunately, in this case the Born-Oppenheimer approximation does not work because the state $\psi_{{\bf R},-}({\bf r})$ is unbound at $R<a$. The problem should then be solved without relying on the adiabatic approximation (see Sec.~\ref{Sec:Aad}).

\section{Role of resonance width}
\label{sec:RoleOfResWidth}

As we have explained in Sec.~\ref{EfimovBO} the Efimov effect is related to the effective $1/R^2$-attraction emerging in a three-body system at distances $R\lesssim |a|$. Let us now discuss how this effective potential is modified in the case of a narrow resonance. First, consider the regime of small detuning, $R_e\ll R^*\ll |a|$. In this regime the right hand side of Eq.~(\ref{eq:kappa}) can be neglected at distances $R^*\ll R\ll |a|$, and we recover the $1/R^2$-behavior of the effective potential (\ref{eq:epsilonShortRWideRes}). However, at distances $R_e\ll R \ll R^*$ the parameter $\kappa_+(R)$ is determined mostly by the last terms on each side of Eq.~(\ref{eq:kappa}). We thus find that $\kappa_+^2(R)\approx 1/R^*R$ and
\begin{equation}\label{BOepsilonNarrowRes}
\epsilon_+(R)\approx -1/2mR^*R,\,\, R_e\ll R \ll R^*\ll |a|.
\end{equation}
The Coulomb potential (\ref{BOepsilonNarrowRes}) is qualitatively different from the inverse-square one (\ref{eq:epsilonShortRWideRes}). In particular, in this potential heavy atoms do not fall to the center. Moreover, close to the origin, $R \ll (m/M)R^*$, the heavy-atom wavefunction behaves similarly to the noninteracting case, i.e. at these distances the kinetic energy operator in the heavy-atom Schr\"odinger equation is dominant. Therefore, we do not have to worry about the boundary condition at the origin (as in Sec.~\ref{sec:RecombinationFermions} we just choose the solution that grows faster). The wavefunction is thus uniquely defined. Then at distances $R\sim R^*$ it can be matched with the Efimov-like wavefunction (\ref{BOCosLog}), the three-body parameter being determined by $R^*$. Remarkably, in this case three-body observables depend only on the two-body parameters $a$ and $R^*$.

It is instructive to discuss more qualitatively what happens with the system as $R$ decreases from $R\gg R^*$ to $R\ll R^*$. One can see that in the former case the effective energy-dependent scattering length $\tilde a(E)$ introduced in Eq.~(\ref{eq:energydependentscatlength}) is larger than $R$ and, therefore, drops out of the problem as in the wide-resonance case with $a=\infty$. However, for $R \sim R^*$ the energy of the light atom becomes sufficiently detuned from the resonance, so that $\tilde a(E)$ is comparable to $R$, and the effective potential acquires a characteristic lengthscale and is no longer that steep. Another important point to mention is a qualitative change of the wavefunction at $R\sim R^*$. By using Eqs.~(\ref{EffRangeClosedChannelProbability}) and (\ref{eq:BOwavefunction}) it is straightforward to show that for $R\gg R^*$ the light atom occupies predominantly the open channel, whereas at distances $R\ll R^*$ the open channel occupation is negligible. Therefore, at these distances the wavefunction of the system to the leading order describes the free motion of a heavy atom and a closed channel molecule. A small open channel occupation can be treated perturbatively.

In the regime of intermediate detuning, $R_e\ll |a| \ll R^*$, the effective potential is nowhere proportional to $1/R^2$. At distances $R\lesssim |a|$ it is approximated by the Coulomb potential (\ref{BOepsilonNarrowRes}). Then in the case $a<0$ there are no bound states for the light atom if $R>|a|$ and the potential $\epsilon_+(R)$ terminates at the point $R=|a|$ where it reaches the three-atom continuum. For positive $a$, the effective potential changes its shape from the Coulomb one for $R\lesssim a$ to the constant $\epsilon_+(R)\approx -|\epsilon_0|$ for $R\gtrsim a$. Accordingly, in both cases trimer states appear only when the characteristic Bohr radius, $\propto (m/M)R^*$, corresponding to the potential (\ref{BOepsilonNarrowRes}) is smaller than $|a|$. 

The last point that we would like to touch upon in this section is inelastic losses in few-body systems near a narrow resonance $R^*$. The approach to this problem is similar to the one discussed in Sec.~\ref{sec:RecombinationFermions} for non-Efimovian systems. In fact, in the narrow-resonance case the probability of relaxation or recombination to deeply bound states can always be related to the relaxation rate constant for collisions of atoms and closed channel molecules. Indeed, the relaxation is a local process, it requires three atoms to approach each other to distances $\sim R_e \ll R^*$ and, as we have argued above, at these distances the wavefunction describes the relative motion of an atom and a closed channel molecule. We will return to this question in Sec.~\ref{Sec:Aad}. More details on the few-body problem near a narrow resonance can be found in
(Petrov 2004; Wang {\it et al.} 2011{\it a}; Levinsen and Petrov 2011).

\chapter{The method of Skorniakov and Ter-Martirosian (STM) for few-body problems with resonant short-range interactions}

\label{Chapter:STM}

In spite of numerous advantages and simplifications provided by the Born-Oppenheimer approximation we still have to discuss methods suitable to quantitatively describe the case of comparable masses and even some highly mass-imbalanced ones (for example, the $s$-wave scattering of a heavy fermion and a weakly bound heavy-light molecule when the symmetry forces the light atom to be in the state $\psi_{{\bf R},-}({\bf r})$ which does not exist at small $R$). 

A natural generalization of the Born-Oppenheimer approximation is the adiabatic hyperspherical method, in which, after separating out the center-of-mass motion, one introduces the coordinate system of hyperradius and hyperangles, $\{\rho,\hat\Omega\}$. The hyperradius $\rho$ is the square root of the sum of squares of all interparticle distances with mass-dependent weights, and all other coordinates of the system can be written as a set of dimensionless hyperangles $\hat\Omega$. The adiabatic idea in this case is to consider $\rho$ formally as a slow coordinate and $\hat\Omega$ as fast ones. Accordingly, one fixes $\rho$ and diagonalizes the hyperangular part of the Schr\"odinger equation thus obtaining the channel potentials. The second step is then to solve the hyperradial part, which is a set of second-order ordinary differential equations for each channel. The difference from the simple Born-Oppenheimer approach is that all these differential equations are now coupled by nonadiabatic matrix elements. In practice one truncates the set of these coupled equations according to a given accuracy goal. Usually only a few channels suffice to obtain convergent results. 

An advantage of the adiabatic hyperspherical approach is that it is quite general. It can be used for finite- and long-range interaction potentials, one can easily include three-body forces, etc. In fact, the complexity of the method is equivalent to the complexity of the initial few-body Schr\"odinger equation and is practically independent of the form of the interatomic potentials. Let us consider an $N$-body problem in three dimensions ($N \ge 3$). Then we can go to the center-of-mass reference frame and separate out the three global Euler angles by using the rotational invariance. We end up with $3N-6$ degrees of freedom, $3N-7$ of which are hyperangular. Thus, for $N=3$, the hyperangular space is two-dimensional, and each additional particle adds 3 degrees of freedom, which makes the hyperangular part of the calculation quite challenging. For the details on the adiabatic hyperspherical approach see, for example, (Lin 1995; Nielsen {\it et al.} 2001)

Another approach, appropriate for systems with short-range interactions, was first introduced by Skorniakov and Ter-Martirosian (Skorniakov and Ter-Martirosian 1957). For $N=3$ it leads to a one-dimensional integral equation in momentum space (the STM equation), and its generalization to $N>3$ gives an integral equation for a function of $3N-9$ coordinates. The STM equation can be obtained by using the Effective Field Theory (EFT) (Bedaque {\it et al.} 1999) and other diagrammatic techniques (Brodsky {\it et al.} 2005; Levinsen and Gurarie 2006). We should mention that the famous Faddeev-Yakubovsky few-body method can be understood as a generalization of the STM approach to the case of finite-range (and long-range) potentials (Faddeev 1960; Yakubovskii 1967; Faddeev and Merkuriev 1993). 

We will now derive the STM equations in coordinate space directly from the Schr\"odinger equation. This derivation clearly shows how one can integrate out all short-range physics and use the fact that mostly everywhere in space the motion of the atoms is free.

\section{STM equation}
\label{SectionSTM}

\begin{figure}[hptb]
\begin{center}
\includegraphics[width=0.5\columnwidth,clip,angle=-90]{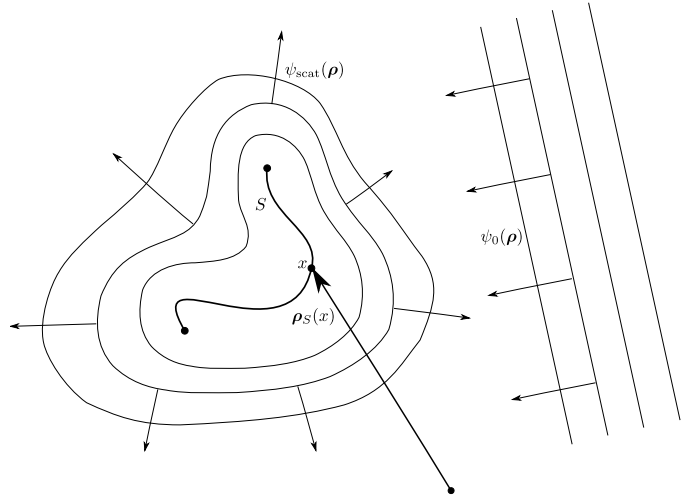}
\caption{Two-dimensional scattering of a particle by a curve.}
\label{fig:ScatByCurve}
\end{center}
\end{figure}

Let us first demonstrate the general idea behind this method. Assume that we are in two-dimensional space and we solve the problem of scattering by a curve $S$ (see Fig.~\ref{fig:ScatByCurve}). Namely, we solve the free Schr\"odinger (or Helmholtz) equation
\begin{equation}\label{eq:Helmholtz2D}
(-\nabla^2_{\mbox{\boldmath$\rho$}}-E)\psi=0
\end{equation}
with the boundary condition
\begin{equation}\label{HelmholtzBoundary}
[\partial\psi/\partial {\bf n}]/\psi=F(x),
\end{equation}  
where $x$ is the natural (by the arc length) parametrization of $S$, $F$ is a known function, and $[\partial\psi/\partial {\bf n}]$ is the sum of the normal outward derivatives at the point $x$ of the curve.

An efficient way of solving this scattering problem is to introduce an auxiliary function $f(x)$ defined on the boundary and look for the solution in the form
\begin{equation}\label{HelmholtzFunction}
\psi({\bf \rho})=\psi_0({\mbox{\boldmath$\rho$}})+\int_S G_E[{\mbox{\boldmath$\rho$}} - {\mbox{\boldmath$\rho$}}_S(x)]f(x){\rm d}x,
\end{equation}  
where $G_E({\mbox{\boldmath$\rho$}})$ is the Green's function of Eq.~(\ref{eq:Helmholtz2D}), ${\mbox{\boldmath$\rho$}}_S(x)$ is the coordinate on the curve, and $\psi_0({\mbox{\boldmath$\rho$}})$ is the incoming wave, which is a free solution of Eq.~(\ref{eq:Helmholtz2D}).

Clearly, the function (\ref{HelmholtzFunction}) satisfies Eq.~(\ref{eq:Helmholtz2D}). We just have to ensure the boundary condition (\ref{HelmholtzBoundary}). It can be shown (by choosing a proper contour around the curve element ${\rm d}x$ and applying the Gauss-Ostrogradsky theorem) that $[\partial\psi/\partial {\bf n}]=f(x)$. Then Eq.~(\ref{HelmholtzBoundary}) gives the equation
\begin{equation}\label{eq.ModelSTMequation}
\psi_0[{\mbox{\boldmath$\rho$}}_S(x)]+\int_S G_E[{\mbox{\boldmath$\rho$}}_S(x) -{\mbox{\boldmath$\rho$}}_S(x')]f(x'){\rm d}x'=1/F(x).
\end{equation}  
Note that Eq.~(\ref{eq.ModelSTMequation}) is one-dimensional in contrast to the original two-dimensional Schr\"odinger equation (\ref{eq:Helmholtz2D}). In fact, it is much more suitable for analytical and especially numerical calculations as this reduction of the configurational space allows for rapid computational schemes. 

The idea of using free Green functions is quite natural when one deals with boundary value problems. It can be encountered in electrostatics, hydrodynamics, problems of heat diffusion, etc. We can mention the so-called method of boundary elements used for calculating spectra of quantum billiards (Berry 1981; Bohigas {\it et al.} 1984) and based on the Korringa-Kohn-Rostoker method (Korringa 1947; Khon and Rostoker 1954) in solid state theory. Let us now discuss how one can do similar things for the few-atom problem.

\begin{figure}[hptb]
\begin{center}
\includegraphics[width=0.25\columnwidth,clip,angle=-90]{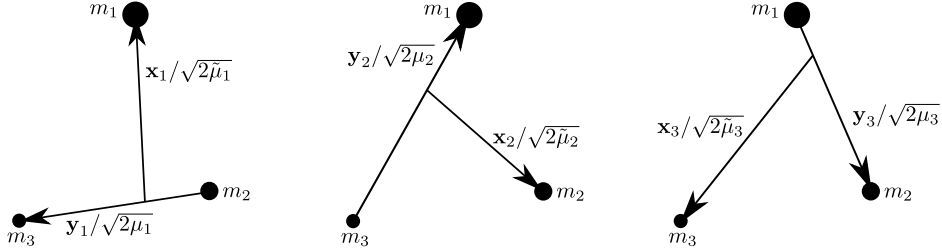}
\caption{Three sets of coordinates for the three-body problem.}
\label{fig:JacobiCoord}
\end{center}
\end{figure}

Consider three atoms with coordinates ${\bf r}_i$ and masses $m_i$, where $i=1,2,3$. We separate out the center of mass motion and introduce the three sets of rescaled Jacobi coordinates (see Fig.~\ref{fig:JacobiCoord}):
\begin{eqnarray}\label{JacobiCoord}
{\bf x}_i&=&\sqrt{2\tilde\mu_i}[{\bf r}_i-(m_j {\bf r}_j+m_k {\bf r}_k)/(m_j+m_k)],\nonumber \\
{\bf y}_i&=&\sqrt{2\mu_i}({\bf r}_k-{\bf r}_j),
\end{eqnarray}  
where $\{i,j,k\}$ are cyclic permutations of $\{1,2,3\}$, i.e. $\{1,2,3\}$, $\{2,3,1\}$, and $\{3,1,2\}$, $\mu_i=m_jm_k/(m_j+m_k)$ is the reduced mass for atoms $j$ and $k$, and $\tilde \mu_i=m_i(m_j+m_k)/(m_i+m_j+m_k)$ is the reduced mass for the relative motion of atom $i$ with respect to pair $jk$. The purpose of the mass-dependent rescaling is to express the kinetic energy operator in the symmetric form $-\nabla^2_{{\bf x}_i}-\nabla^2_{{\bf y}_i}$.

Let us choose to work in the first coordinate system ($i=1$ in Eqs.~(\ref{JacobiCoord})). It is related to the other two by the equations: 
\begin{eqnarray}\label{eq:2-1}
{\bf x}_1={\bf X}_{1\leftarrow 2}({\bf x}_2,{\bf y}_2)&=&-(\sqrt{\mu_1\mu_2}/m_3){\bf x_2}+\sqrt{\mu_2/\tilde\mu_1}{\bf y}_2,\nonumber\\
{\bf y}_1={\bf Y}_{1\leftarrow 2}({\bf x}_2,{\bf y}_2)&=&-\sqrt{\mu_1/\tilde\mu_2}{\bf x_2}-(\sqrt{\mu_1\mu_2}/m_3){\bf y}_2,
\end{eqnarray}
and
\begin{eqnarray}\label{eq:3-1}
{\bf x}_1={\bf X}_{1\leftarrow 3}({\bf x}_3,{\bf y}_3)&=&-(\sqrt{\mu_1\mu_3}/m_2){\bf x_3}-\sqrt{\mu_3/\tilde\mu_1}{\bf y}_3,\nonumber\\
{\bf y}_1={\bf Y}_{1\leftarrow 3}({\bf x}_3,{\bf y}_3)&=&\sqrt{\mu_1/\tilde\mu_3}{\bf x_3}-(\sqrt{\mu_1\mu_3}/m_2){\bf y}_3.
\end{eqnarray}

Now the original three-body problem can be represented as a single particle scattering in the six-dimensional space by a potential which is localized on the three-dimensional hyperplanes defined by the equations ${\bf y}_i=0$, $i=1,2,3$. Everywhere outside these hyperplanes the three-body wavefunction satisfies the free-motion six-dimensional Schr\"odinger equation
\begin{equation}\label{eq:Helmholtz}
(-\nabla^2_{{\bf x}_1}-\nabla^2_{{\bf y}_1}-E)\Psi({\bf x}_1,{\bf y}_1)=0.
\end{equation}
As in the usual scattering theory we write $\Psi({\bf x}_1,{\bf y}_1)=\Psi_0({\bf x}_1,{\bf y}_1)+\Psi_{\rm scat}({\bf x}_1,{\bf y}_1)$, where $\Psi_0({\bf x}_1,{\bf y}_1)$ is the incoming wave -- a free solution of Eq.~(\ref{eq:Helmholtz}) without singularities (as if there were no interactions), and $\Psi_{\rm scat}({\bf x}_1,{\bf y}_1)$ is the scattered wave which contains $1/y$-singularities at each of the boundaries. We now introduce auxiliary functions defined at the boundaries, $f_i({\bf x}_i)$, and write $\Psi({\bf x}_1,{\bf y}_1)$ in the form
\begin{eqnarray}\label{PsiThroughf}
&&\hspace{-1.1cm}\Psi({\bf x}_1,{\bf y}_1)=\Psi_0({\bf x}_1,{\bf y}_1) + \int G_E(\sqrt{({\bf x}_1-{\bf x'})^2+y_1^2})f_1({\bf x'}){\rm d}^3 x'\nonumber\\
&&\hspace{-0.8cm}+\sum_{i=2,3}\int G_E(\sqrt{[{\bf x}_1-{\bf X}_{1\leftarrow i}({\bf x'},0)]^2+[{\bf y}_1-{\bf Y}_{1\leftarrow i}({\bf x'},0)]^2})]f_i({\bf x'}){\rm d}^3 x',
\end{eqnarray} 
where $G_E$ is the Green's function of Eq.~(\ref{eq:Helmholtz}):
\renewcommand{\arraystretch}{2} 
\begin{equation}\label{eq:6DGreen}
G_E(X)=\left\{\begin{array}{lr}\frac{-EK_2(\sqrt{-E}\left|X\right|)}{8\pi^3\!
X^2},&E<0\\
\frac{iEH_2(\sqrt{E}\left|X\right|)}{16\pi^2
X^2},&E>0\end{array}\right.
\xrightarrow[E\rightarrow 0]{}
\frac{1}{4\pi^3 X^4}.
\end{equation}
\renewcommand{\arraystretch}{1}

By construction, the wavefunction (\ref{PsiThroughf}) satisfies the Schr\"odinger equation (\ref{eq:Helmholtz}) for arbitrary boundary functions $f_i({\bf x}_i)$. As in the previous example this freedom is removed by boundary conditions. We first discuss the wide resonance case. Having in mind the Bethe-Peierls boundary condition (\ref{BP2}) let us study the singular and regular terms of $\Psi({\bf x}_1,{\bf y}_1)$ close to the boundary $y_1\rightarrow 0$. In this limit $\Psi_0({\bf x}_1,{\bf y}_1)$ and the two terms in the sum in the second line of Eq.~(\ref{PsiThroughf}) are not singular. We can thus set $y_1=0$ there. In the remaining term we make the $1/y_1$-singularity explicit by subtracting and adding the quantity 
\begin{equation}\label{eq:AddSubtract}
f_1({\bf x}_1) \int G_E(\sqrt{({\bf x}_1-{\bf x'})^2+y_1^2}){\rm d}^3 x'=f_1({\bf x}_1)e^{-\sqrt{-E}y_1}/4\pi y_1,
\end{equation}
where, for $E>0$ we use the convention $\sqrt{-E}=-i\sqrt{E}$. We thus get the following asymptotic expression for $\Psi({\bf x}_1,{\bf y}_1)$:
\begin{equation}\label{eq:PsiAsymptote}
\Psi({\bf x}_1,{\bf y}_1)\xrightarrow[y_1\rightarrow 0]{} (1/4\pi) [(1/y_1 -\sqrt{-E})f_1({\bf x}_1) +4\pi\Psi_0({\bf x}_1,0) - \hat{L}_{1,E}\{f\}({\bf x}_1)],
\end{equation}
where we have expanded the right hand side of Eq.~(\ref{eq:AddSubtract}) to the next-to-leading order at small $y_1$, $\{f\}=\{f_1,f_2,f_3\}$, and $\hat{L}_{1,E}$ is the integral operator defined by
\begin{eqnarray}\label{eq:Loperator}
&&\hspace{-1.1cm}\hat{L}_{1,E}\{f\}({\bf x}_1)=4\pi\int G_E(|{\bf x}_1-{\bf x'}|)[f_1({\bf x}_1)-f_1({\bf x'})]{\rm d}^3 x'\nonumber\\
&&\hspace{-0.8cm}-4\pi\sum_{i=2,3}\int G_E[\sqrt{x_1^2+x'^2+2\sqrt{\mu_1/\mu_i}(1-\mu_i/m_1){\bf x}_1 {\bf x'}}]f_i({\bf x'}){\rm d}^3 x'.
\end{eqnarray} 

Comparing Eqs.~(\ref{eq:PsiAsymptote}) and (\ref{BP2}) we arrive at the integral equation
\begin{equation}\label{eq:IntEqWideRes}
\hat{L}_{1,E}\{f\}({\bf x}_1)+(\sqrt{-E}-1/\sqrt{2\mu_1}a_1)f_1({\bf x}_1)=4\pi \Psi_0({\bf x}_1,0),
\end{equation}
where $a_1$ is the scattering length corresponding to the interaction of atoms 2 and 3.

Applying the same procedure to the boundaries ${\bf y}_2=0$ and ${\bf y}_3=0$ we can derive two additional equations, the left hand sides of which can be obtained from that of Eq.~(\ref{eq:IntEqWideRes}) by cyclically permuting subscripts 1,2, and 3. The right hand sides of the new equations read, respectively, $4\pi \Psi_0[{\bf X}_{1\leftarrow 2}({\bf x}_2,0),{\bf Y}_{1\leftarrow 2}({\bf x}_2,0)]$ and $4\pi \Psi_0[{\bf X}_{1\leftarrow 3}({\bf x}_3,0),{\bf Y}_{1\leftarrow 3}({\bf x}_3,0)]$. We thus get three coupled integral equations for determining the boundary functions $f_i$ and we have substantially reduced the configurational space of the problem. In fact, the operators $\hat{L}_{i,E}$ conserve angular momentum, and we can expand $f_i$ in spherical harmonics to work only with a set of one-dimensional integral equations. Solution of these equations then gives us the wavefunction $\Psi$ by virtue of Eq.~(\ref{PsiThroughf}). Vice versa, if we know $\Psi$, we can always obtain $f_i$ by looking at the coefficients in front of the $1/y_i$-singularities of $\Psi$ at small $y_i$. Namely,
\begin{equation}\label{f}
f_i({\bf x}_i)=4\pi \lim_{y_i \rightarrow 0}y_i \Psi.
\end{equation}
Thus, $\{f\}$ contains the same information about the system as the wavefunction $\Psi$. This is one of the key advantages of the zero-range approximation. 

It is sometimes useful to go to momentum representation introducing the Fourier transform $f_k({\bf p})=\int f_k({\bf x})\exp (-i{\bf p}{\bf x}){\rm d}^3 x$. Then the operator $\hat{L}_{1,E}\{f\}$ reads
\begin{eqnarray}\label{eq:LoperatorMomentum}
&&\hspace{-1.1cm}\hat{L}_{1,E}\{f\}({\bf p}_1)=(\sqrt{-E+p_1^2}-\sqrt{-E})f_1({\bf p}_1)\nonumber\\
&&\hspace{-0.8cm}-\frac{1}{2\pi^2}\sum_{i=2,3}\sqrt{\frac{\tilde\mu_1}{\mu_i}} \int \frac{f_i({\bf p'})\,{\rm d}^3p'}{p'^2+p_1^2+2\sqrt{\mu_1/\mu_i}(1-\mu_i/m_1){\bf p}_1{\bf p'}-(\mu_i/\tilde\mu_1)E}.
\end{eqnarray} 
Skorniakov and Ter-Martirosian derived Eq.~(\ref{eq:IntEqWideRes}) in momentum representation in the case of equal masses in order to solve the problem of neutron-deutron scattering (Skorniakov and Ter-Martirosian 1957). We will refer to Eq.~(\ref{eq:IntEqWideRes}) as the STM equation. Let us now illustrate the power of the method by considering some concrete examples.

\section{The Efimov effect and determination of $s_0$}
\label{Sec:Exacts0}

We return to the problem of Sec.~\ref{sec:QuantumStatistics} and show how to determine the parameter $s_0$ without relying on the Born-Oppenheimer approximation. The system we are interested in consists of two identical atoms (bosons or fermions) of mass $m_1=m_2=M$ interacting resonantly (scattering length $a$) with another atom of mass $m_3=m$. Let us neglect the (nonresonant) interaction between the identical atoms. Then $\Psi$ is not singular for $y_3\rightarrow 0$ and therefore $f_3\equiv 0$. By symmetry the boundary functions $f_1$ and $f_2$ are equal to each other (bosons) or have different signs (fermions). Indeed, directly from Eq.~(\ref{JacobiCoord}) we see that the permutation ${\bf r}_1\rightleftarrows {\bf r}_2$ is equivalent to ${\bf x}_1\rightleftarrows {\bf x}_2$ and ${\bf y}_1\rightleftarrows -{\bf y}_2$. Thus $\Psi({\bf x}_1,{\bf y}_1)=\pm\Psi({\bf x}_2,-{\bf y}_2)$ and by virtue of Eq.~(\ref{f}) we have $f({\bf x}):=f_1({\bf x})=\pm f_2({\bf x})$, where the upper sign stands for bosons and lower -- for fermions. Clearly, $\Psi_0$ should have the same property. Finally, the three equations of type (\ref{eq:IntEqWideRes}) are reduced in this case to the single one,
\begin{equation}\label{eq:IntEqXXY}
\hat{L}_E f({\bf x})+(\sqrt{-E}-1/\sqrt{2\mu}a)f({\bf x})=4\pi \Psi_0({\bf x},0),
\end{equation}  
where 
\begin{equation}\label{eq:LoperatorXXY}
\hat{L}_E f({\bf x})=4\pi\int \{G_E(|{\bf x}-{\bf x'}|)[f({\bf x})-f({\bf x'})]\mp G_E(\sqrt{x^2+x'^2+2{\bf x} {\bf x'}\sin\phi })f({\bf x'})\}{\rm d}^3 x',
\end{equation} 
$\mu=\mu_1=\mu_2=mM/(m+M)$, and $\sin\phi=M/(m+M)$.

At small $x \ll \{a, \sqrt{|E|}\}$ we can neglect all terms in Eq.~(\ref{eq:IntEqXXY}) except the integral one and in the latter we can use the zero-energy asymptote of the Green's function (\ref{eq:6DGreen}). This means that the small-$x$ asymptote of $f({\bf x})$ should satisfy
\begin{equation}\label{eq:STMZeroEnergyXXY}
\hat{L}_{E=0} f({\bf x})=0.
\end{equation}
We first consider spherically symmetric $f({\bf x})=f(x)$. In this case averaging over the angles of ${\bf x}'$ (we integrate over the solid angle and divide by $4\pi$) we obtain the zero momentum component of the operator $\hat{L}_{E=0}$:
\begin{equation}\label{eq:L0XXY}
\hat L_{E=0,l=0} f(x)=\frac{4}{\pi} \int_0^\infty \left[\frac{f(x)-f(x')}{(x^2-x'^2)^2}\mp \frac{f(x')}{(x^2+x'^2)^2-4x^2x'^2\sin^2 \phi}\right] x'^2 d x'.
\end{equation} 
The first integral in Eq.~(\ref{eq:L0XXY}) [and also in Eqs.~(\ref{eq:Loperator}) and (\ref{eq:LoperatorXXY})] is taken in the principal value sense (Petrov {\it et al.} 2005{\it a}). We see that the operator (\ref{eq:L0XXY}) is scaleless and has the property 
\begin{equation}\label{eq:propertyXXY}
\hat L_{E=0,l=0} x^\nu =\lambda_{l=0}(\nu) x^{\nu-1}.
\end{equation}
We thus look for the solution of Eq.~(\ref{eq:STMZeroEnergyXXY}) in the form $f(x)=x^\nu$, where $\nu$ is a root of $\lambda_{l=0}(\nu)$. The region of convergence of the integral on the left hand side of Eq.~(\ref{eq:propertyXXY}) is $-3<{\rm Re}(\nu)<1$ in the case of bosons and $-5<{\rm Re}(\nu)<3$ for fermions. The function $\lambda_{l=0}(\nu)$ is given by
\begin{equation}\label{eq:lambdaSwaveXXY}
\lambda_{l=0}(\nu)=-(\nu + 1)\tan\frac{\pi\nu}{2}\mp\frac{2\sin[\phi(\nu+1)]}{\sin(2\phi)\cos(\pi\nu/2)}.
\end{equation}
The integrals of type $\hat L_{E=0,l=0} x^\nu$ can be taken by using the complex analysis. We cut the complex $x$-plane along the positive real axis and choose the integration contour embracing this cut. Then the contour can be blown to infinity, and the integral is determined by the pole residues of the integrand. 

\begin{figure}[hptb]
\begin{center}
\includegraphics[width=0.6\columnwidth,clip,angle=0]{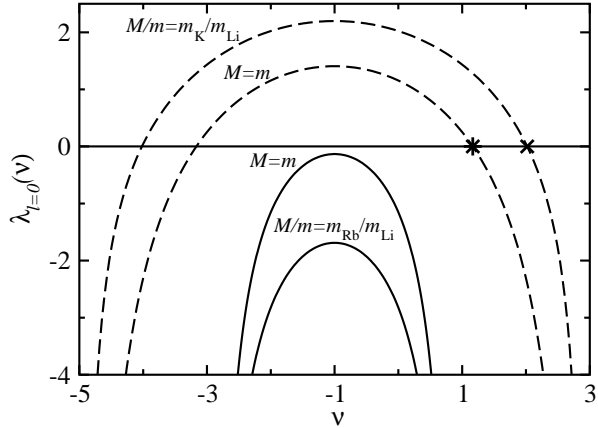}
\caption{$\lambda_{l=0}(\nu)$ versus $\nu$ in the case of two identical bosons (solid lines) or two identical fermions (dashed lines) interacting resonantly with a third atom of mass $m$. The Efimov effect takes place only for bosons. The star and cross denote the value $\nu_1\approx 1.16622$ for equal masses and $\nu_1\approx 2.0193$ for the case of two $^{40}K$ and one $^6$Li atoms.}
\label{fig:lambdaSwave}
\end{center}
\end{figure}

In Fig.~\ref{fig:lambdaSwave} we plot $\lambda_{l=0}(\nu)$ in the bosonic (solid lines) and fermionic (dashed lines) cases. In the fermionic case for any mass ratio the equation $\lambda(\nu)=0$ has two solutions $\nu_1>\nu_2$ in the region of convergence mentioned above. These roots are real which means that we are in the non-Efimovian case discussed in Sec.~\ref{sec:RecombinationFermions}. Accordingly, in the absence of a three-body resonance $f(x)\propto x^{\nu_1}$. Note that we have just solved the problem that could not be solved in the Born-Oppenheimer approximation even at very large mass ratios. Indeed, in the case of heavy identical fermions the symmetry $l=0$ corresponds to the light atom in the state $\psi_-$, which is unbound at these small distances.

The case of identical bosons in the Born-Oppenheimer language corresponds to the wavefunction $\psi_+$ for the light atom and zero total angular momentum (see Sec.~\ref{sec:QuantumStatistics}). In this case the Born-Oppenheimer approach predicts the Efimov effect. Now, studying $\lambda_{l=0}(\nu)$ one can show that the effect takes place for any mass ratio as the roots of this function are complex conjugate, $\nu_{1,2}=-1\pm is_0$, where $s_0$ is real. Accordingly, the three-body parameter is necessary in order to fix the ratio in between the coefficients in the linear superposition of $x^{\nu_1}$ and $x^{\nu_2}$. The parameter $s_0$ plotted in Fig.~\ref{Fig:s0} (solid line) is found by solving the equation $\lambda_{l=0}(-1+is_0)=0$. The increase of $s_0$ with the mass ratio is consistent with the Born-Oppenheimer picture.

As a historical remark we note that the neutron-deutron $s$-wave scattering problem considered by Skorniakov and Ter-Martirosian for total spin $I=3/2$ corresponds to the fermionic case discussed above and for total spin $I=1/2$ -- to the bosonic case. Indeed, in the latter case the neutron spins are antiparallel and their orbital wavefunction is symmetric as for bosons. It was later pointed out by Danilov (Danilov 1961) that for $I=1/2$ the solution of the STM equation (\ref{eq:IntEqXXY}) is not unique and an additional parameter is necessary. 

We should also note that the bosonic case with equal masses is characterized by relatively small $s_0\approx 0.414$ leading to very large scaling factor $\exp(\pi/s_0)\approx 1986.1$. This fact, although not disturbing for theorists, causes apparent practical difficulties. The reason for this ``weak'' manifestation of the Efimov physics is that there are only two resonant interactions out of three. However, as we have already mentioned in Sec~\ref{sec:QuantumStatistics}, the scaling factor rapidly decreases with the mass ratio (see upper panel of Fig.~\ref{Fig:s0}), which makes highly mass-imbalanced heteronuclear mixtures practically valuable for studies of the Efimov effect and discrete scaling invariance. 

The case $l=1$ is treated in the same manner as $l=0$. We obtain the operator $\hat L_{E=0,l=1}$ by integrating out the angular dependence of $f({\bf x})=3f(x)\hat{\bf x}\hat{\bf n}$ in $\hat L_{E=0}f({\bf x})$ (we multiply it by $\hat{\bf x}\hat{\bf n}$, integrate over the solid angle, and divide by $4\pi$):
\begin{eqnarray}\label{eq:L1XXY}
&&\hspace{-1.1cm}\hat L_{E=0,l=1} f(x)=\frac{1}{\pi} \int_0^\infty \left\{\frac{4x'^2[f(x)-f(x')]}{(x'^2-x^2)^2}-\frac{f(x')}{x^2}\left[\frac{2xx'}{(x+x')^2}-\log\frac{x+x'}{|x-x'|}\right]\right.\nonumber\\
&&\hspace{-0.8cm}\mp \left.\frac{f(x')}{2x^2\sin^2\phi}\left[\frac{4xx'(x^2+x'^2)\sin\phi}{x^4+x'^4+2x^2x'^2\cos(2\phi)}+\log\frac{x^2+x'^2-2xx'\sin\phi}{x^2+x'^2+2xx'\sin\phi}\right]\right\} d x'.
\end{eqnarray} 
The operator (\ref{eq:L1XXY}) has the same scaling property (\ref{eq:propertyXXY}) with 
\begin{equation}\label{eq:lambdaPwaveXXY}
\lambda_{l=1}(\nu)=\frac{\nu(\nu + 2)}{\nu+1}\cot\frac{\pi\nu}{2}\mp\frac{\nu\sin(\phi)\cos[(\nu+1)\phi]-\sin(\nu\phi)}{(\nu+1)\sin^2(\phi)\cos(\phi)\sin(\pi\nu/2)}.
\end{equation}
The region of convergence of the integrals in Eq.~(\ref{eq:L1XXY}) is $-4<{\rm Re}(\nu)<2$ for bosonic and for fermionic symmetries.

\begin{figure}[hptb]
\begin{center}
\includegraphics[width=0.6\columnwidth,clip,angle=0]{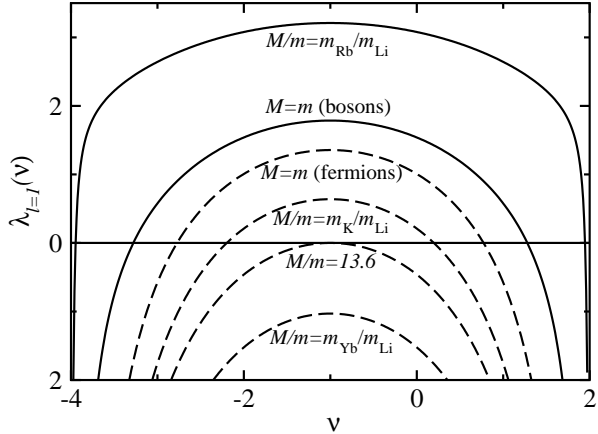}
\caption{$\lambda_{l=1}(\nu)$ versus $\nu$ in the case of two identical bosons (solid lines) or two identical fermions (dashed lines) of mass $M$ interacting resonantly with a third atom of mass $m$. For this symmetry ($l=1$) the Efimov effect is absent in the bosonic case. For fermions it takes place only for mass ratios larger than the critical one, $(M/m)_c \approx 13.6$.}
\label{fig:lambdaPwave}
\end{center}
\end{figure}

In Fig.~\ref{fig:lambdaPwave} we plot the function $\lambda_{l=1}(\nu)$ for bosons and for fermions. In the Born-Oppenheimer language this symmetry ($l=1$) corresponds to the antisymmetric heavy-atom wavefunction. In the bosonic case, besides leading to the centrifugal barrier, it also forces the light atom to be in state $\psi_-$ (leading to an effective repulsion for the heavy atoms). This explains the absence of the Efimov effect for bosons in this case. For fermions we have the competition of the centrifugal barrier and the attraction due to the exchange of the light atom in state $\psi_+$. For mass ratios $M/m>13.6$ the latter wins and the Efimov effect manifests itself in the complex roots of $\lambda_{l=1}(\nu)$, which have the same structure as in the case of zero angular momentum, $\nu_{1,2}=-1\pm is_0$. The quantity $s_0$ for the $l=1$ case is also plotted in Fig.~\ref{Fig:s0} along with the result of the Born-Oppenheimer approximation.

\section{Analytical approach at zero total energy}
\label{sec:analytics}

In this section we would like to present an approach which allows one to calculate all zero-energy three-body observables analytically. The approach was introduced for the fermionic non-Efimovian case in (Petrov 2003) and has been recently generalized to Efimovian cases in heteronuclear mixtures (Helfrich {\it et al.} 2010). Here, we apply it to three identical bosons. Alternative analytical methods of treating this system can be found in (Macek {\it et al.} 2005; Macek and Ovchinnikov 2006; Gogolin {\it et al.} 2008; Mora {\it et al.} 2011{\it a}). We should mention that our current understanding of the three-boson problem is strongly influenced by the papers of Nielsen and Macek (1999), Esry {\it et al.} (1999), Bedaque {\it et al.} (2000), and Braaten and Hammer (2001), who analyzed the process of three-body recombination in this system numerically and arrived at a number of important conclusions. 

Close to a Feshbach resonance a Bose gas suffers from three-body recombination -- formation of weakly (for large positive $a$) or deeply bound molecules, the binding energy being transfered to the kinetic energy of the products -- a molecule and remaining atom. Let $\alpha_s$, $\alpha_d(a>0)$, and $\alpha_d(a<0)$ stand for the rate constants for recombination to a weakly bound state and deeply bound states on the positive and negative sides of the resonance, respectively. In the low temperature limit, as long as the de Broglie wavelengths are larger than $|a|$, these constants are practically temperature independent and can be written as products of $\hbar a^4/m$ and dimensionless log-periodic functions of $|a|/R_0$ [different for $\alpha_s$, $\alpha_d(a>0)$, and $\alpha_d(a<0)$]. For fixing the phase of these log-periodic functions the following practical notation is used in literature: the value of $a>0$ where $\alpha_s/a^4$ reaches its minimum is denoted by $a_{*0}$ [defined modulo $\exp(\pi/s_0)$]. Another experimentally relevant reference point is the maximum of $\alpha_d(a<0)/a^4$, which is by definition reached at $a=a_-<0$. In fact, this is the point where a Efimov state crosses the three-atom threshold. We will show below that $|a_-|/a_{*0}=\exp(\pi/2s_0)$.   

Let us first discuss the case $a>0$ and temporarily adopt the units $\hbar=m=a=1$. For the problem of three-body recombination we now consider Eq.~(\ref{eq:IntEqWideRes}) and choose the incoming wave $\Psi_0$ as a symmetrized superposition of plain waves normalized to a volume $V$. In the region of space relevant for recombination it can be approximated by
\begin{equation}\label{Psi_0}
\Psi_0=\sqrt{6}V^{-3/2},
\end{equation}
where the factor $\sqrt{6}$ is due to the fact that we consider cold bosons but not in the same quantum state (not condensed). Let us mention that Eq.~(\ref{eq:IntEqWideRes}) with $E=0$ can also be used for the problem of atom-dimer scattering just below the breakup threshold ($E=-0$). In this case the atoms cannot move freely at large distances and $\Psi_0$ should be set to zero.

Using the bosonic symmetry and Eq.~(\ref{f}) it is straightforward to show that $f({\bf x}):=f_1({\bf x})=f_2({\bf x})=f_3({\bf x})$ (we work in the first coordinate system, $\{ {\bf x}_1,{\bf y}_1\}$, and omit the subscript). Equation~(\ref{eq:IntEqWideRes}) is thus a single three-dimensional integral equation. Moreover, we can set $f({\bf x})=f(x)$ consistent with the angular independence of $\Psi_0$ and with the fact that all processes with higher angular momenta are suppressed at low temperatures. The resulting STM equation reads:
\begin{equation}\label{CoorSTM}
(\hat L_{E=0,l=0} -1)f(x)=4\pi \Psi_0,
\end{equation} 
where the integral operator $\hat L_{E=0,l=1}$ is obtained in the same manner as (\ref{eq:L0XXY}). Omitting the subscripts, we get
\begin{equation}\label{L0}
\hat L f(x)=\frac{4}{\pi} \int_0^\infty \left[\frac{f(x)-f(x')}{(x^2-x'^2)^2}-\frac{2f(x')}{(x^2+x'^2)^2-x^2x'^2}\right] x'^2 d x'.
\end{equation} 

Let us now discuss the structure of possible solutions of Eq.~(\ref{CoorSTM}). Obviously, $f(x)$ is a sum of a particular  solution of the inhomogeneous equation (\ref{CoorSTM}) and a general solution of the homogeneous equation
\begin{equation}\label{STMHom}
(\hat L -1)\chi(x)=0.
\end{equation}
Physically, Eq.~(\ref{STMHom}) describes the atom-dimer channel just below the dimer breakup threshold ($\Psi_0 = 0$). Therefore, at distances $x\gg 1$ the function $\chi(x)$ is a linear combination of $\exp(ix)/x$ and $\exp(-ix)/x$. Indeed, consider the atom-dimer wavefunction $\Psi(x,y)=\phi_b(y)\exp(\pm ix)/x$. Substituting it into Eq.~(\ref{f}) and using the correctly normalized dimer wavefunction 
\begin{equation}\label{phi_b}
\phi_b(y)=\exp(-y)/\sqrt{2\pi}y\xrightarrow[y\rightarrow 0]{}(1/y-1)/\sqrt{2\pi}
\end{equation}
we find that the corresponding contribution to $\chi(x)$ equals $(8\pi)^{1/2}\exp(ix)/x$. 

Our aim now is to solve Eq.~(\ref{CoorSTM}), separate the large-$x$ asymptote $f\propto \exp(ix)/x$, and relate the coefficient in front of it to the three-body recombination rate constant. This problem can be solved analytically by using the property (\ref{eq:propertyXXY}) of the operator $\hat L$. In the case of three identical bosons the integral converges in the region $-3<{\rm Re}(\nu)<1$ and the function $\lambda_{l=0}(\nu)$ is given by (we again omit the subscript)
\begin{equation}\label{lambda}
\lambda(\nu)=-(\nu + 1)\tan\frac{\pi\nu}{2}-\frac{8}{\sqrt{3}}\frac{\sin[\pi(\nu+1)/6]}{\cos(\pi\nu/2)}.
\end{equation}
The roots of (\ref{lambda}) are complex conjugate, $\nu_{1,2}=-1\pm is_0$, where $s_0$ satisfies
\begin{equation}\label{EqForS0Bosons}
s_0\cosh (\pi s_0/2)-8\sinh (\pi s_0/6)/\sqrt{3}=0.
\end{equation}
The solution is $s_0\approx 1.00624$ which leads to the famous scaling factor $\exp (\pi/s_0)\approx 22.7$.

At short distances the operator $\hat L$ in Eq.~(\ref{STMHom}) dominates over 1, and any solution of this equation should 
be a linear superposition of $\chi\propto x^{-1+is_0}$ and its complex conjugate. From now on we will use the notation $\chi$ 
for the solution of Eq.~(\ref{STMHom}) with the following asymptotes:
\renewcommand{\arraystretch}{1} 
\begin{equation}\label{chi}
\chi(x)=\left\{\begin{array}{lr}Ax^{\nu}=Ax^{-1+is_0},&x\ll1,\\
x^{-1}e^{ix+i\sigma-h}+x^{-1}e^{-ix-i\sigma+h},&x\gg1,
\end{array}\right.
\end{equation}
\renewcommand{\arraystretch}{1}where $A$ is a complex number, and $\sigma$ and $h$ are real numbers. The physical solution of 
Eq.~(\ref{STMHom}), i.e. the one corresponding to a given three-body parameter, is expressed as
\begin{equation}\label{chi_physical}
\chi_{\theta}(x)=e^{i\theta}\chi(x)+e^{-i\theta}\chi^*(x),
\end{equation}
where we have introduced the three-body parameter $\theta$ (a complex number with imaginary part $\eta_*$).

The normalization in Eq.~(\ref{chi}) is chosen such that
\begin{equation}\label{orthonorm}
\langle p\chi(px)|p'\chi(p'x)\rangle\!=\!\!\int_0^\infty\!\!\! p\chi(px) p'\chi(p'x) x^2 dx=2\pi\delta (p-p').
\end{equation} 
The first equality in Eq.~(\ref{orthonorm}) is our definition of the scalar product (note the absence of the complex conjugation), 
and the second equality follows from the fact that $p\chi(px)$ and $p'\chi(p'x)$ are eigenfunctions of the symmetric operator 
$\hat L$ corresponding to the eigenvalues $p$ and $p'$. They are orthogonal for $p\neq p'$ and their scalar product in the 
vicinity of $p=p'$ can be worked out in the same way as in (Landau and Lifshitz 1987, see \S 21). A simple change of the 
integration variable in Eq.~(\ref{orthonorm}) leads to the completeness condition
\begin{equation}\label{completeness}
\int_0^\infty p^2\chi(px)\chi(px')dp=2\pi\delta (x-x')/x^2.
\end{equation}

Equations~(\ref{orthonorm}) and (\ref{completeness}) allow us to construct the integral operator $(\hat L-1)^{-1}$ needed 
to solve Eq.~(\ref{CoorSTM}). In order to avoid problems with divergence of the corresponding integrals let us introduce 
an auxiliary function $g_0(x)$ related to $f(x)$ by
\begin{equation}\label{ftog_0}
f(x)=4\pi\Psi_0 [-1-\lambda(0)/x+\lambda(0)\lambda(-1)g_0(x)].
\end{equation}
Substituting this expression into Eq.~(\ref{CoorSTM}) and using Eq.~(\ref{eq:propertyXXY}) we find that $g_0(x)$ satisfies 
the equation $(\hat L-1)g_0(x)=x^{-2}$. Applying the operator $(\hat L-1)^{-1}$ to $x^{-2}$ we obtain the following particular solution
\begin{equation}\label{partsolution}
g_0(x)=\frac{1}{2\pi x}\int_0^\infty \chi(z)dz \left[\int_0^\infty \frac{\chi(y)ydy}{y-x-i0}-\frac{2\pi i x\chi(x)}{1-\exp(-2\pi s_0)}\right ],
\end{equation}
where the first integral is defined as
\begin{equation}\label{sense}
\int_0^\infty \chi(z) dz=\lim_{\epsilon\rightarrow +0}\int_0^\infty \chi(z)z^\epsilon dz.
\end{equation}
The rule of going around the pole in the second integral and the numerical coefficient in front of the second term in the 
square brackets on the right hand side of Eq.~(\ref{partsolution}) regulate the entry of $\chi(x)$, which can be arbitrary, 
into the particular solution $g_0(x)$. Using this freedom we choose these parameters in such a way that $g_0(x)$ does not 
contain oscillating terms proportional to $x^{-1+is_0}$ at small $x$. Direct calculation shows that in the limit 
$x\rightarrow 0$ the right hand side of Eq.~(\ref{partsolution}) equals $g_0(x)\approx [\int_0^\infty\chi(z)dz]^2/2\pi x$ 
to the leading order in $x$. On the other hand, according to Eq.~(\ref{eq:propertyXXY}) the same quantity in the same limit can 
be written as $g_0(x)=(\hat L -1)^{-1}x^{-2}\approx \hat L^{-1}x^{-2}=1/[\lambda(-1)x]$, which leads to the result
\begin{equation}\label{intchi}
\int_0^\infty\chi(z)dz=\sqrt{2\pi/\lambda(-1)}.
\end{equation}
Another consequence of our choice of the particular solution (\ref{partsolution}) is that removing the oscillating terms 
from $g_0(x)$ makes it real, since any imaginary part of $g_0$ would necessarily be a solution of the homogeneous 
Eq.~(\ref{STMHom}). Therefore, $g_0$ would have oscillations at short $x$, the absence of which we have ensured. 
Clearly, the function $f$ obtained by virtue of Eq.~(\ref{ftog_0}) is also real. Moreover, the property (\ref{eq:propertyXXY}) 
ensures that $f=o(1)$ at small $x$, i.e. its Taylor expansion starts with $x^1$, at least. Therefore, this solution of 
Eq.~(\ref{CoorSTM}) is not sensitive to the short-range physics and does not depend on the three-body parameter. 

Integrating Eq.~(\ref{partsolution}) in the limit $x\gg 1$ we get
\begin{equation}\label{g0largex}
g_0(x)\xrightarrow[x\rightarrow \infty]{}\frac{-i}{\sinh(\pi s_0)}\sqrt{\frac{2\pi}{\lambda(-1)}}\frac{\cos[x+\sigma+i(h+\pi s_0)]}{x}.
\end{equation}
It can be real only if $h=-\pi s_0$ (note that $\lambda(-1)<0$), cf. (Macek {\it et al.} 2005). 

\begin{figure}[hptb]
\begin{center}
\includegraphics[width=0.6\columnwidth,clip,angle=0]{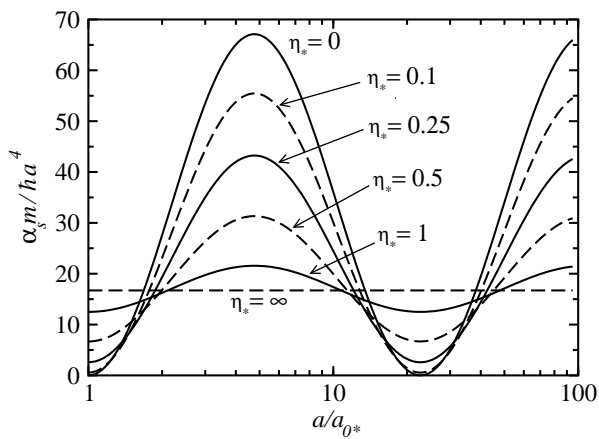}
\caption{The rate constant $\alpha_s$ for the recombination into a weakly bound state for three identical bosons. The function is log-periodic with the scaling factor $\exp(\pi/s_0)\approx 22.7$.}
\label{Fig:AlphaS}
\end{center}
\end{figure}

Finally, the result that we are interested in is the linear combination
\begin{equation}\label{f_theta}
f_\theta (x)=f(x)+\gamma \chi_\theta (x),
\end{equation}
where the complex number $\gamma$ is chosen such that $f_\theta (x)$ contains only an outgoing wave at large $x$ 
(this corresponds to an atom and a dimer flying apart after the three-body recombination event). This condition gives
\begin{equation}\label{beta}
\gamma=i\frac{\pi\Psi_0\lambda(0)\sqrt{2\pi\lambda(-1)}}{\sinh(\pi s_0)\cosh(\pi s_0-i\theta)}.
\end{equation}
Keeping only the relevant oscillating term at large $x$ we obtain
\begin{equation}\label{amplitude}
f_\theta (x)\xrightarrow[x\rightarrow \infty]{} i 4\gamma  \sin\theta \sinh(\pi s_0)\exp(ix+i\sigma)/x.
\end{equation}
So, we have found the coefficient in front of the outgoing atom-dimer wave, which is enough to calculate the atom-dimer outgoing flux. Indeed, the large-$x$ asymptote $f=\xi \exp(ix)/x$, where $\xi$ is any complex amplitude, is accompanied by the flux $|\xi|^2\Phi_\infty$, where
\begin{equation}\label{Phi_infty}
\Phi_\infty=3\times (8\pi)^{-1} \times (4\pi) \times 2 = 3.
\end{equation}
Here we have explicitly written out the following factors: the factor of 3 reflects the three symmetric possibilities of forming 
the dimer (corresponds to the interchange ${\bf r_1}\rightleftarrows {\bf r_2}\rightleftarrows {\bf r_3}$), the factor of $(8\pi)^{-1}$ arises from the 
relation in between $\Psi$ and $f$ (see the discussion after Eq.~(\ref{STMHom})), the factor of $4\pi$ is the solid angle 
in the outgoing atom-dimer channel, and the last factor of 2 is the atom-dimer relative velocity in the $x,y$-coordinates. 
The three-body recombination rate constant $\alpha_s$ is obtained by taking the squared modulus of the prefactor in front of 
$\exp(ix+i\sigma)/x$ in Eq.~(\ref{amplitude}) and by multiplying it by $\Phi_\infty$, by the factor of $1/6$ reflecting the 
fact that the number of triples in the gas is $n^3/3!$, and by the factor $\hbar a^4/m$ in order to restore 
the original physical units. We should also mention that the 9-dimensional volume $V^3$ is taken to be a unit volume in the original system of coordinates $\{{\bf r}_1,{\bf r}_2,{\bf r}_3\}$. In the new coordinates $\{{\bf x}, {\bf y}, {\bf R}_{\rm cm}\}$, 
where ${\bf R}_{\rm cm}$ is the center-of-mass coordinate, this volume equals $V^3=8/3\sqrt{3}$. The final result for the 
three-body recombination rate constant reads
\begin{equation}\label{eq:alphasexact}
\alpha_s= 128\pi^2(4\pi-3\sqrt{3})\frac{\sin^2[s_0\ln(a/a_{*0})]+\sinh^2\eta_*}{\sinh^2(\pi s_0+\eta_*)+\cos^2[s_0\ln(a/a_{0*})]}\frac{\hbar a^4}{m},
\end{equation}
where we have expressed the three-body parameter $\theta$ through the original physical units:
\begin{equation}\label{theta}
\theta=s_0\ln(a/a_{*0})+i\eta_*.
\end{equation}
In Fig.~\ref{Fig:AlphaS} we plot the quantity $m\alpha_s/\hbar a^4$ as a function of $a$ for different values of the elasticity parameter $\eta_*$.

The rate of recombination into a weakly bound level is only a fraction of the total loss rate. The rest is due to the formation of deeply bound molecular states. We have already discussed this mechanism in Sec.~\ref{sec:RecombinationBosons}. In contrast to the recombination into shallow states we now 
have to look at the balance of the incoming and outgoing fluxes of atoms corresponding to the short distance asymptote of 
$f_\theta (x)$ given by Eq.~(\ref{f_theta})
\begin{equation}\label{amplitude_0}
f_\theta (x) \xrightarrow[x\rightarrow 0]{} \gamma \chi_\theta (x)=\gamma (A e^{i\theta}x^{-1+is_0}+A^* e^{-i\theta}x^{-1-is_0}),
\end{equation} 
In analogy with $\Phi_\infty$ let $\Phi_0$ denote the number of atom triples disappearing at the origin (${\bf x}=0$, ${\bf y}=0$), 
provided the function $f$ takes the form of the incoming wave $x^{-1-is_0}$ with unit weight. With this definition the 
recombination rate constant follows from Eq.~(\ref{amplitude_0}):
\begin{equation}\label{alpha_d}
\alpha_d=(1/3!)(\hbar a^4/m) |\gamma|^2 \Phi_0 |A|^2 2\sinh(2\eta_*),
\end{equation}
where the prefactor $1/3!$ is related to the number of triples in the gas. The product $\Phi_0 |A|^2$ can easily be found from the definition (\ref{chi}) by equating the fluxes at $x\rightarrow 0$ and at $x\rightarrow \infty$ and using Eq.~(\ref{Phi_infty}):
\begin{equation}\label{fluxes}
\Phi_0 |A|^2=2 \Phi_\infty \sinh(2\pi s_0) =6\sinh(2\pi s_0).
\end{equation}
Substituting Eqs.~(\ref{fluxes}) and (\ref{beta}) into Eq.~(\ref{alpha_d}) we obtain
\begin{equation}\label{eq:alphadexact}
\alpha_d(a>0)= 128\pi^2(4\pi-3\sqrt{3})\frac{\coth(\pi s_0)\cosh(\eta_*)\sinh(\eta_*)}{\sinh^2(\pi s_0+\eta_*)+\cos^2[s_0\ln(a/a_{0*})]}\frac{\hbar a^4}{m}.
\end{equation} 
Since in the case of identical bosons the product $\pi s_0$ is rather large the function $\alpha_d(a>0)/a^4$ is almost flat (we can neglect the $a$-dependent term in the denominator) in contrast to $\alpha_s/a^4$. The constant $\alpha_d(a>0)$ monotonically increases with $\eta_*$ and in the extreme limit $\eta_*\rightarrow \infty$ the ratio between the two rate constants equals $\alpha_d(a>0)/\alpha_s=\coth(\pi s_0)\approx 1.0036$.

\begin{figure}[hptb]
\begin{center}
\includegraphics[width=0.6\columnwidth,clip,angle=0]{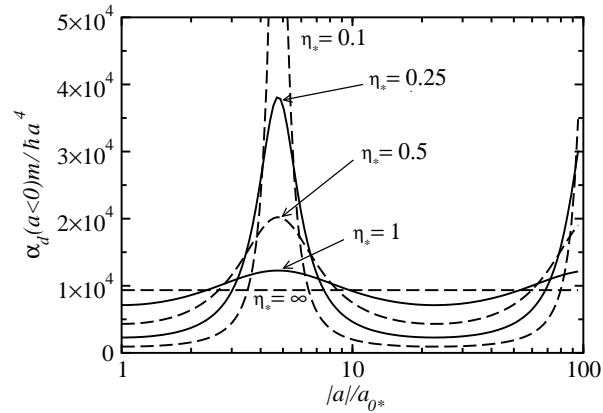}
\caption{The rate constant $\alpha_d$ for the recombination into deeply bound states for three identical bosons on the negative side of the Feshbach resonance. The peaks correspond to Efimov states crossing the three-atom threshold.}
\label{Fig:AlphaD}
\end{center}
\end{figure}

Let us now discuss the negative side of the resonance and derive $\alpha_d(a<0)$. In this case  we use the units $\hbar=|a|=m=1$ and some equations described above should be modified accordingly. In particular, Eq.~(\ref{CoorSTM}) reads
\begin{equation}\label{CoorSTMnega}
(\hat L +1)\tilde f(x)=4\pi \Psi_0
\end{equation} 
and we now introduce an auxiliary function $\tilde g_0$ related to $\tilde f$ by
\begin{equation}\label{ftog_0nega}
\tilde f(x)=4\pi\Psi_0 [1-\lambda(0)/x+\lambda(0)\lambda(-1)\tilde g_0(x)],
\end{equation}
where $\tilde g_0$ satisfies $(\hat L +1)\tilde g_0(x)=x^{-2}$. We write the solution in the form
\begin{equation}\label{partsolutionnega}
\tilde g_0(x)=\frac{1}{2\pi x}\int_0^\infty \chi(z)dz \int_0^\infty \frac{\chi(y)ydy}{y+x}
\end{equation}
and integrating it in the small-$x$ limit we get the asymptote
\begin{equation}\label{fsmallxnega}
\tilde f(x)\xrightarrow[x\rightarrow 0]{} i \frac{2\pi \Psi_0 \lambda(0)\sqrt{2\pi \lambda(-1)}}{\sinh(\pi s_0)}Ax^{-1+is_0}.
\end{equation}
The function $\tilde f$ is a solution of Eq.~(\ref{CoorSTMnega}), but its oscillations at small $x$ do not have (in general) 
the correct phase imposed by Eq.~(\ref{chi_physical}). This difficulty is resolved by observing that $\tilde f^*$ also satisfies 
Eq.~(\ref{CoorSTMnega}). The correctly behaving solution reads
\begin{equation}\label{f_thetanega}
\tilde f_\theta(x)=\frac{\exp(i\theta)\tilde f(x)+\exp(-i\theta)\tilde f^*(x)}{\exp(i\theta)+\exp(-i\theta)},
\end{equation}
and by subtracting the outgoing flux from the incoming one at small $x$ we obtain the result
\begin{equation}\label{alpha_dnega}
\alpha_d(a<0)=128\pi^2(4\pi-3\sqrt{3})\frac{\coth(\pi s_0)\cosh(\eta_*)\sinh(\eta_*)}{\cos^2[s_0\ln(|a|/a_{0*})]+\sinh^2(\eta_*)}\frac{\hbar a^4}{m}.
\end{equation}
For small $\eta_*$ Eq.~(\ref{alpha_dnega}) is characterized by resonances at $a=-a_{0*}\exp(\pi/2s_0)$ modulo $\exp(\pi/s_0)$ (see Fig.~\ref{Fig:AlphaD}). These points mark the passage of Efimov trimers across the three-atom threshold with the usual consequences: the contribution of the trimer state (of size $\sim |a|$) in the three-body wavefunction becomes very large (close to $\theta=\pi/2$ the right hand side of Eq.~(\ref{f_thetanega}) diverges) and atoms spend a lot of time close to each other, which leads to enhanced recombination losses. Note that on average $\alpha_d(a<0)$ is significantly larger (by about three orders of magnitude) than $\alpha_s$ and $\alpha_d(a>0)$ for the same $|a|$. In the limit $\eta_*\rightarrow \infty$ we have $\alpha_d(a<0)/\alpha_s=\exp(2\pi s_0)\coth(\pi s_0)\approx 558.9$. 

Recalling that $a_-$ is defined as the value of $a$ where $\alpha_d(a<0)/a^4$ reaches its maximum, Eq.~(\ref{alpha_dnega}) implies 
\begin{equation}\label{analyticratio}
|a_-|/a_{0*}=\exp(\pi/2s_0).
\end{equation}
The zero-range theory thus predicts that the maxima of $\alpha_s$ and $\alpha_d(a>0)$ and the maxima of $\alpha_d(a<0)$ are placed at $a=\pm |a_-|$, i.e. symmetric with respect to the center of the Feshbach resonance.

This analytical approach can be extended to homonuclear and heteronuclear mixtures of fermions and bosons by taking into account the mass imbalance and/or nonzero angular momentum. It can be used as a zero-energy reference point for numerical approaches which are supposed to give results also at finite energies (three-body recombination at finite temperatures, atom-dimer scattering, etc.) It is also of immense help in the cases of small $s_0$ where the exponentially large scaling parameter impedes numerical calculations.

\section{STM equation near a narrow resonance}
\label{STMNarrowRes}

Let us now discuss how one can include the finite width of the resonance into the STM approach. In fact, the derivation of Sec.~\ref{SectionSTM} is also valid in the narrow resonance case except that the STM equation itself (\ref{eq:IntEqWideRes}) should be modified. Indeed, the asymptote (\ref{eq:PsiAsymptote}) should be compared with the narrow resonance Bethe-Peierls condition (\ref{BP4}) which in current notations reads
\begin{equation}\label{BP6}
\Psi({\bf x}_1,{\bf y}_1)\propto 1/y_1 -1/\sqrt{2\mu_1}a_1-\sqrt{2\mu_1}R_1^* E_1,\; y_1 \rightarrow 0,
\end{equation}
where $E_1$ is the collision energy of atoms 2 and 3, i.e. the kinetic energy of motion ``along'' the ${\bf y}_1$-direction of the six-dimensional $\{{\bf x}_1,{\bf y}_1\}$-space. Using Eq.~(\ref{eq:Helmholtz}) and the definition (\ref{f}) we write
\begin{equation}\label{eq:CollisionEnergy}
E_1=\lim_{{\bf y}_1\rightarrow 0}\frac{-\nabla_{{\bf y}_1}^2\Psi}{\Psi}=E+\lim_{{\bf y}_1\rightarrow 0}\frac{\nabla_{{\bf x}_1}^2\Psi}{\Psi}=E+\frac{\nabla_{{\bf x}_1}^2f_1}{f_1}.
\end{equation}
Substituting Eq.~(\ref{eq:CollisionEnergy}) into Eq.~(\ref{BP6}) and comparing the latter with (\ref{eq:PsiAsymptote}) we obtain the narrow resonance STM equation
\begin{equation}\label{eq:IntEqNarrowRes}
\hat{L}_{1,E}\{f\}({\bf x}_1)+[\sqrt{-E}-1/\sqrt{2\mu_1}a_1-\sqrt{2\mu_1}R_1^* (E+\nabla_{{\bf x}_1}^2)]f_1({\bf x}_1)=4\pi \Psi_0({\bf x}_1,0).
\end{equation}
The equations for $f_2$ and $f_3$ should also be modified by including similar terms proportional to $R_2^*$ and $R_3^*$, respectively. Equation~(\ref{eq:IntEqNarrowRes}) can be, of course, written in momentum space. The operator $-\nabla_{{\bf x}_1}^2$ is then substituted by $p_1^2$ and the corresponding representation of $\hat{L}_{1,E}$ is given by Eq.~(\ref{eq:LoperatorMomentum}).

At sufficiently short distances the operator $\nabla_{{\bf x}_1}^2$ in Eq.~(\ref{eq:IntEqNarrowRes}) dominates and the short-distance asymptote of $f_1$ satisfies $-\nabla_{{\bf x}_1}^2 f_1({\bf x}_1)=0$. This is consistent with Sec.~\ref{sec:RoleOfResWidth} where we argued that at these distances the three-body wavefunction describes the free motion of an atom and a closed channel molecule. In fact, according to Eq.~(\ref{EffRangeClosedChannelProbability}) the probability for atoms 2 and 3 to be in the closed channel equals $4\pi R^*_1|\lim_{y_1\rightarrow 0}y_1\Psi|^2=4\pi R^*_1 |f_1({\bf x}_1)/4\pi|^2$, and, therefore, $\sqrt{R^*_1/4\pi}f_1({\bf x}_1)$ can be considered as the wavefunction of the relative motion of an atom and a bare molecule. In the narrow resonance limit they interact only by virtually breaking up the molecule and exchanging one of its constituents with the free atom. This exchange interaction is given by the integral operator $\hat{L}_{1,E}$ in Eq.~(\ref{eq:IntEqNarrowRes}) and can be treated perturbatively in the limit $R^*\rightarrow \infty$. Obviously, we can also introduce the direct interaction between the closed channel molecule and atom by imposing another zero-range Bethe-Peierls boundary condition, this time on the function $f_1$. 

\section{Atom-dimer scattering near a narrow resonance}
\label{Sec:Aad}  

To illustrate how the approach of Sec.~\ref{STMNarrowRes} can be implemented in practice let us consider a concrete example. Namely, we calculate the atom-dimer scattering length for a system of two statistically identical atoms of mass $M$ and another atom of mass $m$. We use the notations of Sec.~\ref{Sec:Exacts0} with upper sign for identical bosons and lower -- for fermions. To account for the finite resonance width we now add the term $-\sqrt{2\mu}R^* (E+\nabla_{{\bf x}}^2)f({\bf x})$ to the left hand side of Eq.~(\ref{eq:IntEqXXY}) and look for its solution at zero atom-molecule collision energy, i.e. we set $E=\epsilon_0<0$, where $\epsilon_0$ is the energy of the molecular state [see Eq.~(\ref{BoundStateEnergy})]. Since $E$ is negative, the free solution $\Psi_0$ vanishes, and Eq.~(\ref{eq:IntEqXXY}) takes the form
\begin{equation}\label{eq:IntEqXXYAtomDimer}
(-\sqrt{2\mu}R^*\nabla_{{\bf x}}^2+\hat{L}_{E=\epsilon_0})f({\bf x})=0.
\end{equation}

The algorithm of calculating the atom-dimer scattering length $a_{ad}$ is straightforward: expand Eq.~(\ref{eq:IntEqXXYAtomDimer}) in spherical harmonics, solve it for $l=0$ ($s$-wave symmetry), and deduce $a_{ad}$ from the large-$x$ asymptote of $f(x)$. Indeed, at distances $x\gg 1/\kappa$, where $\kappa=\sqrt{2\mu |\epsilon_0|}$, Eq.~(\ref{eq:IntEqXXYAtomDimer}) describes the free atom-molecule motion, i.e. the $s$-wave symmetric solution is a linear superposition of $x^{-1}$ and $x^0$. The relation between the corresponding coefficients is fixed by the atom-dimer scattering length:
\begin{equation}\label{eq:AtomDimerLargeX}
f(x) \xrightarrow[x\rightarrow \infty]{} N (1-\sqrt{2\tilde\mu}a_{ad}/x),
\end{equation}
where $\tilde \mu=m(M+m)/(2M+m)$ is the atom-molecule reduced mass and $N$ is the normalization prefactor, which is actually not needed for determining $a_{ad}$, but we will return to it when discussing the inelastic atom-dimer relaxation. 

Alternatively, one can follow the same procedure in momentum space. Introducing the Fourier transform $f({\bf p})=\int f({\bf x})\exp (-i{\bf p}{\bf x}){\rm d}^3 x$ we look for the solution in the form
\begin{equation}\label{fthroughg}
f(p)=N[(2\pi)^3\delta({\bf p})+4\pi g(p/\sqrt{-\epsilon_0})/(\sqrt{-\epsilon_0}p^2)].
\end{equation}
We substitute (\ref{fthroughg}) into the momentum space STM equation, integrate over the angles of ${\bf p}$ and  arrive at the following equation for the function $g(k)$,
\begin{eqnarray}\label{eq:EqForGMomentum}
&&\hspace{-1.1cm}\sqrt{-2\mu \epsilon_0}R^*g(k)=\pm\frac{1}{(k^2+\cos^2\phi)\cos\phi}\nonumber\\
&&\hspace{-1.1cm}-\frac{\sqrt{1+k^2}-1}{k^2} g(k)\pm\frac{1}{\pi\sin (2\phi)}\int_0^\infty \frac{g(k')}{kk'}\log\frac{k^2+k'^2+2kk'\sin\phi+\cos^2\phi}{k^2+k'^2-2kk'\sin\phi+\cos^2\phi}{\rm d}k',
\end{eqnarray} 
the solution of which gives the atom-dimer scattering length: $a_{ad}=-g(0)/\sqrt{-2\tilde \mu \epsilon_0}$.

\begin{figure}[hptb]
\begin{center}
\includegraphics[width=0.6\columnwidth,clip,angle=0]{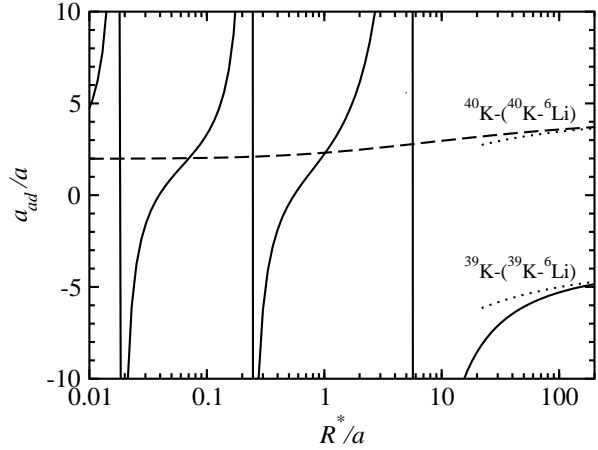}
\caption{The atom-dimer $s$-wave scattering length for $^{39}$K-($^{39}$K-$^6$Li) (solid line) and $^{40}$K-($^{40}$K-$^6$Li) (dashed line) collisions near a narrow interspecies resonance. The dotted lines show the large $R^*/a$ asymptotes (\ref{TwoTermsAad}).}
\label{Fig:Aad}
\end{center}
\end{figure}

In Fig.~\ref{Fig:Aad} we plot $a_{ad}$ in units of $a$ calculated from Eq.~(\ref{eq:EqForGMomentum}) for collisions of K atoms with K-Li molecules formed near a narrow interspecies resonance. The solid line shows the case of identical bosonic $^{39}$K atoms and the dashed one -- identical fermionic $^{40}$K. We see that in the limit $R^*/a\ll 1$ the bosonic case is Efimovian and is characterized by the log-periodic dependence of $a_{ad}/a$ on $a$ of the form (\ref{BOAtomDimerScatLength}), the parameter $R^*$ plays the role of the three-body parameter as we have discussed in Sec.~\ref{sec:RoleOfResWidth}. In contrast, in the fermionic case the atom-dimer scattering length has a well-defined limit for $R^*=0$ and the left hand side of Eq.~(\ref{eq:EqForGMomentum}) can be considered as a weak perturbation for small $R^*/a$.

In the opposite limit, $R^*/a\gg 1$, which we call the regime of intermediate detuning, we treat Eq.~(\ref{eq:EqForGMomentum}) perturbatively in the small parameter $\eta=1/\sqrt{-2\mu \epsilon_0}R^*\ll 1$. Namely, we look for the solution in the form $g(k)=g^{(0)}(k)+g^{(1)}(k)+g^{(2)}(k)+...$, where the functions $g^{(i)}$ are of order $\eta^{i}$ and can be found recursively: $g^{(0)}\equiv 0$, $g^{(1)}(k)= \pm \eta \cos^{-1}(\phi)/(k^2+\cos^2\phi)$, then $g^{(i)}$ for $i>1$ is obtained by substituting $g^{(i-1)}$ into the $g$-dependent terms in the right hand side of Eq.~(\ref{eq:EqForGMomentum}) and by multiplying the result by $\eta$. In this way we obtain the first two leading terms
\begin{equation}\label{TwoTermsG}
g(0)\approx \pm\frac{\eta}{\cos^3\phi}\left[1+\frac{-1\pm\cos^{-2}\phi}{2}\eta\right].
\end{equation}
Then, expressing the small parameter in terms of $\sqrt{a/R^*}$ and using
\begin{equation}\label{TwoTermsEta}
\eta=\frac{2}{\sqrt{1+4R^*/a}-1}\approx\sqrt{\frac{a}{R^*}}\left(1+\frac{1}{2}\sqrt{\frac{a}{R^*}}\right) 
\end{equation}
we finally obtain the atom-dimer scattering length up to the next-to-leading order in $\sqrt{a/R^*}$:
\begin{equation}\label{TwoTermsAad}
\frac{a_{ad}}{a}\approx\mp\frac{1}{\cos^2\phi}\left(1+\frac{1\pm \cos^{-2}\phi}{2}\sqrt{\frac{a}{R^*}}\right). 
\end{equation}
The dotted lines in Fig.~\ref{Fig:Aad} show the asymptotes (\ref{TwoTermsAad}).

Let us now discuss how the atom-dimer relaxation can be calculated for molecules formed near a narrow resonance. The process is local, it happens at distances of the order of the size of the closed channel molecule, which is much smaller than $R^*$ according to our definition of narrow resonances. We have already mentioned that the probability of finding three-atoms in a small volume of size $\ll R^*$ is dominated by the probability of finding there an atom and a closed channel molecule. Therefore, the probability density for finding three atoms in the recombination region equals $(R^*/4\pi)|f(x=0)|^2$, which is proportional to the relaxation rate if we treat the process perturbatively. The proportionality prefactor requires solving the three-body problem at short distances, which is a challenging task. However, since the shape of the closed channel wavefunction is not dramatically sensitive to the magnetic field, we can assume that this microscopic prefactor stays approximately constant close to a given Feshbach resonance. Therefore, the quantity that can be calculated in the zero-range approximation is the ratio $\alpha_{\rm rel}(a)/\alpha_{\rm bare}$ of the relaxation rate constant at a given $a$ to the relaxation rate constant for collisions of atoms and closed channel molecules. This ratio should tend to 1 in the limit $R^*\gg a$.

\begin{figure}[hptb]
\begin{center}
\includegraphics[width=0.6\columnwidth,clip,angle=0]{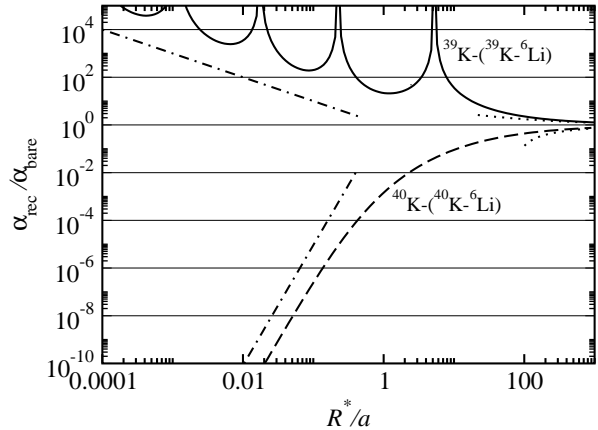}
\caption{The atom-dimer $s$-wave relaxation rate constant versus the detuning. The physical systems and notations are the same as in Fig.~\ref{Fig:Aad}. The dash-dotted lines reflect the power scaling of $\alpha_{\rm rel}$ at large $a$ (see text).}
\label{Fig:Alphaad}
\end{center}
\end{figure}

In Fig.~\ref{Fig:Alphaad} we plot the ratio $\alpha_{\rm rel}(a)/\alpha_{\rm bare}$ for the same physical systems and using the same notations as in Fig.~\ref{Fig:Aad}, i.e. the solid line stands for bosonic $^{39}$K and dashed -- for fermionic $^{40}K$. The results are obtained by substituting the already calculated function $g(k)$ into Eq.~(\ref{fthroughg}) and evaluating $f(x=0)$ by integrating $f(k)$ over momenta. One should keep track of the $a$-dependence of the normalization prefactor $N\propto \alpha_{\rm norm}$ given by Eq.~(\ref{NormalizationAlpha}). This dependence follows from the correct normalization of the dimer wavefunction in the narrow resonance case (see Sec.~\ref{MoleculeNarrowRes}). The final result reads
\begin{equation}\label{Eq:Alphaad}
\frac{\alpha_{\rm rel}(a)}{\alpha_{\rm bare}}=\frac{\sqrt{1+4R^*/a}-1}{\sqrt{1+4R^*/a}}\left|1+\frac{2}{\pi}\int_0^\infty g(k){\rm d}k\right|^2.
\end{equation}
The dotted lines in Fig.~\ref{Fig:Alphaad} show the large $R^*/a$-asymptotes and are obtained by approximating $g(k)\approx g^{(1)}(k)$ in Eq.~(\ref{Eq:Alphaad}).

In the regime of small detuning, $R^*/a\ll 1$, the dependence of $\alpha_{\rm rel}/\alpha_{\rm bare}$ on $R^*/a$ can be estimated qualitatively. Consider first the fermionic case. The function $f(x)$ is of order $N\sim 1/\sqrt{a}$ at distances $\sim a$ where it should be matched with the power law $f(x)\sim a^{-1/2}(x/a)^{\nu_1}$ approximately valid in the window $R^*\lesssim x\lesssim a$. The power $\nu_1$ is the root of $\lambda_{l=0}(\nu)$ [see Eq.~(\ref{eq:lambdaSwaveXXY}) and Fig.~\ref{fig:lambdaSwave}]. For the system $^{40}$K-$^{40}$K-$^6$Li it equals $\nu_1\approx 2.0193$. Then at distances $x\lesssim R^*$ the function $f$ is approximately constant since the interaction in between the atom and the closed channel molecule is not resonant. We finally get
\begin{equation}\label{AlphaadFermiScaling}
\alpha_{\rm rel}/\alpha_{\rm bare}\propto R^*|f(x=0)|^2\propto (R^*/a)^{-2\nu_1-1},\,\, R^*\ll a.
\end{equation}
This qualitative scaling is shown in Fig.~\ref{Fig:Alphaad} as the lower dash-dotted line. 

We see that the $s$-wave atom-dimer relaxation is highly suppressed in the limit $R^*\ll a$. A good news is that even for $a=R^*$, where the qualitative scaling (\ref{AlphaadFermiScaling}) shows no suppression, the quantitative result predicts the suppression of the relaxation by three orders of magnitude compared to the ``bare'' atom-molecule case. Apparently, for this relatively large mass ratio the exchange of the light atom leads to an appreciable repulsion between the heavy ones (remember that according to the Born-Oppenheimer approach, the light atom is in the ``repulsive'' $\psi_-$ state). These effects are not taken into account in the derivation of Eq.~(\ref{AlphaadFermiScaling}).  

As far as the bosonic case is concerned we can repeat the same speculations disregarding the atom-dimer resonances and the oscillatory part of $f(x)\sim a^{-1/2}x^{-1}\cos[s_0\ln (x/r_0)]$ at distances $R^*\lesssim x\lesssim a$. We obtain then for bosons $\alpha_{\rm rel}/\alpha_{\rm bare}\sim (R^*/a)^{-1}$, i.e. Eq.~(\ref{AlphaadFermiScaling}) with $\nu_1=-1$ (the upper dash-dotted line in Fig.~\ref{Fig:Alphaad}). Figure~\ref{Fig:Alphaad} clearly shows that in this case there is no suppression of the relaxation. 

\chapter{Final remarks}
\section{Outlook}

Let us now briefly mention other rapidly developing directions of the theoretical few-atom physics. Molecular regimes in fermionic and bosonic mixtures require a better understanding of collisional properties of molecules, which is a four-atom problem. The transition from three-body to four-body is conceptually straightforward, it requires just a bit more space for formulas, although much more computing power. The four-body STM equation for fermions in coordinate and momentum space has been derived along the lines of Chap.~\ref{Chapter:STM} by Petrov {\it et al.} (2004, 2005{\it a}, 2005{\it b}). The three-body STM approach has also been generalized to the four-body case by using diagrammatic techniques (Brodsky {\it et al.} 2005; Levinsen and Gurarie 2006). The Born-Oppenheimer approximation has been applied for calculating collisional properties of highly mass imbalanced heteronuclear molecules (Marcelis {\it et al.} 2008). Various aspects of the four-boson problem with short-range interactions (universality, tetramer states, necessity of an additional four-body parameter) have been actively studied in momentum space by using the Faddeev-Yakubovsky equations (Platter {\it et al.} 2004{\it a}; Yamashita {\it et al.} 2006, Hammer and Platter 2007; Deltuva 2010), by using the adiabatic hyperspherical approach (von Stecher {\it et al.} 2009; see also review of Rittenhouse {\it et al.} 2011), and by other methods (Hanna and Blume 2006; von Stecher 2010; Yamashita {\it et al.} 2010). The problem of three heavy fermions interacting with a light one has recently attracted attention (Castin {\it et al.} 2010; Blume and Daily 2010; Gandolfi and Carlson 2010; Mora {\it et al.} 2011{\it b}). In particular, Castin {\it et al.} (2010) have argued that in this system the four-body Efimov effect can occur for mass ratios $M/m<13.6$, i.e. in the absence of the three-body one.

Another direction of research, obviously relevant for ultracold gases, is the few-body problem in a trap. The spectrum of three trapped bosons has been calculated by Jonsell {\it et al.} (2002). The relation of the trapped problem at unitarity ($a=\infty$) and the free-space problem at zero energy has been established by Tan (2004) and by Werner and Castin (2006). Various results on the spectrum and correlation functions of few-fermion systems in a harmonic potential have been obtained by Blume and co-workers (von Stecher {\it et al.} 2007, 2008; Blume {\it et al.} 2007). For an overview of the few-body problem in a trap see (Blume 2012). 

A notable progress has been made in solving the few-body problem in low dimensions. The hyperspherical approach has been applied to the two-dimensional three-boson problem by Nielsen {\it et al.} (1999) and by Kartavtsev and Malykh (2006). Platter {\it et al.} (2004{\it b}) have calculated the bound state energies of four two-dimensional bosons by using the Faddeev-Yakubovsky equations. Brodsky {\it et al.} (2005) have considered the two-dimensional few-body problem in the diagrammatic framework. Heteronuclear two-dimensional trimers have been discussed by Pricoupenko and Pedri (2010) by using the two-dimensional STM equation in momentum space. Liu {\it et al.} (2010{\it b}) have calculated the second and third virial coefficients for a two-dimensional strongly interacting Fermi gas.

As far as the one-dimensional case is concerned the problem of N particles is exactly solvable in some cases (Lieb and Liniger 1963; Lieb 1963; McGuire 1965; Yang 1967; Gaudin 1967) and the corresponding wavefunctions can be written out explicitly. The exact results can be compared to the results of various few-body methods (Dodd 1970; Thacker 1975; Amaya-Tapia {\it et al.} 2004). Speaking of nonintegrable cases, the atom-dimer and dimer-dimer scattering problem in quasi-one-dimensional geometry has been studied by Mora and co-workers (Mora {\it et al.} 2004, 2005{\it a}, 2005{\it b}). Muryshev {\it et al.} (2002), Sinha {\it et al.} (2006), and Mazets {\it et al.} (2008) have perturbatively estimated the three-body diffraction effect originating from the finite transversal size in a quasi-1D system of bosons. Deviations from integrability originating from the finite width of a Feshbach resonance have been investigated by Yurovsky {\it et al.} (2006). The spectral statistics and the response of 1D few-body systems with and without mass imbalance (nonintegrable and integrable, respectively) have been studied by Colom\'e-Tatch\'e and Petrov (2011). The one-dimensional three-body problem on a lattice has recently been discussed by Orso {\it et al.} (2011) and Valiente {\it et al.} (2010), see also (Keilmann {\it et al.} 2009).

Different components of a heteronuclear mixture can feel different external forces. For example, we can imagine that one of the components is confined to a quasi-one-dimensional or quasi-two-dimensional geometry whereas the other component remains free. Besides, by adding an optical lattice one can separately control the effective masses of the components. This leads to interesting peculiarities of many-, few-, and even two-body problems in such mixed-dimensional systems (Petrov {\it et al.} 2007; Nishida and Tan 2008; Levinsen {\it et al.} 2009; Lamporesi {\it et al.} 2010).

We should mention the analysis of the few-fermion problem near a $p$-wave resonance (Levinsen {\it et al.} 2007; Jona-Lasinio {\it et al.} 2008) which is interesting in view of the realization of nontrivial superfluid phases in strongly interacting polarized fermionic gases. Finally, the progress in preparing cold dipolar gases (see the lecture course of D.~S. Jin and J. Ye in this book) motivates studies of the few-body problem with dipolar interactions (Wang {\it et al.} 2011{\it b}, 2011{\it c}).

\section{Acknowledgements}

I would like to thank the organizers of this summer school, Gora Shlyapnikov and Christophe Salomon, for giving me the opportunity to teach at Les Houches. I also thank them as my collaborators on the topic of these lectures. I am very much indebted to people with whom I collaborated or deeply discussed the few-body physics, Maria Colom\'e, Kerstin Helfrich, Jesper Levinsen, Hans-Werner Hammer, Tobias Tiecke, Jook Walraven, Bout Marcelis, Servaas Kokkelmans, Gregory Astrakharchik, David Papoular, Victor Gurarie, Maxim Efremov, Mikhail Baranov, Felix Werner, Yvan Castin, Ludovic Pricoupenko, Maxim Olshanii, Vladimir Yurovsky, Doerte Blume, Dmitri Fedorov, Mikhail Zvonarev, Pietro Massignan, Roman Krems, and many others. The work on these lectures was financially supported by the EuroQUAM-FerMix program, by the ANR grant BLAN0165, and by the IFRAF Institute.

\end{document}